\documentclass[]{aa}  

\usepackage{graphicx}
\usepackage{txfonts}

\usepackage{float}
\usepackage{caption}
\usepackage{subcaption}
\usepackage{xcolor}

\begin{document}

   \title{{\tt SubDLe}:  identification of substructures in cosmological \\ simulations with deep learning}

   \subtitle{An image segmentation approach to substructure finding}

   \author{Michela Esposito\inst{\ref{inst1},\ref{inst2},\ref{inst3},\ref{inst4}} 
          \and
          Stefano Borgani\inst{\ref{inst1},\ref{inst2},\ref{inst3},\ref{inst4},\ref{inst5}}
          \and
          Giuseppe Murante\inst{\ref{inst2},\ref{inst3},\ref{inst5}}
          }

   \institute{Department of Physics, Astronomy Section, University of Trieste, via G. Tiepolo 11, I-34131 Trieste, Italy\\
              \email{michela.esposito@phd.units.it}\label{inst1}
         \and
             INAF – Osservatorio Astronomico di Trieste, via G. Tiepolo 11, I-34143 Trieste, Italy\label{inst2}\\
             \email{michela.esposito@inaf.it}
        \and
            IFPU – Institute for Fundamental Physics of the Universe, Via Beirut 2, I-34014 Trieste, Italy\label{inst3}
        \and
            INFN – Instituto Nazionale di Fisica Nucleare, Via Valerio 2, I-34127, Trieste, Italy \label{inst4}
        \and
            ICSC - Italian Research Center on High Performance Computing, Big Data and Quantum Computing, via Magnanelli 2, 40033, Casalecchio di Reno, Italy\label{inst5}
             }

   \date{Received MMM DDD, YYY; accepted MMM DDD, YYY}

 
  \abstract
   {The identification of substructures within halos in cosmological hydrodynamical simulations is a fundamental step to identify the simulated counterparts of real objects, namely galaxies. For this reason, substructure finders play a crucial role in extracting relevant information from the simulation outputs. In general, they are based on physically-motivated definitions of substructures, performing multiple steps of particle-by-particle operations, and for this reason they are computationally expensive.}
   {The purpose of this work is to develop a fast algorithm to identify substructures, especially galaxies, in simulations. The final aim, besides a faster production of subhalo catalogues, is to provide an algorithm fast enough to be applied with a fine time-cadence during the evolution of the simulations. Having access to galaxy catalogues while the simulation is evolving is necessary indeed for sub-resolution models based on global properties of galaxies.}
   {In this context, Machine Learning methods offer a wide range of automated tools for fast analysis of large data sets. So, we chose to apply the architecture of a well known Fully Convolutional Network, the {\tt U-Net}, to the identification of substructures within the mass density field of the simulation. We have developed {\tt SubDLe} ({\tt Sub}structure identification with {\tt D}eep {\tt Le}arning), an algorithm which combines a 3D generalization of {\tt U-Net} and a Friends-of-Friends algorithm, and trained it to reproduce the identification of substructures performed by the {\tt SubFind} algorithm in a set of zoom-in cosmological hydrodynamical simulations of galaxy clusters. For the feasibility study presented in this work, we have trained and tested {\tt SubDLe} on galaxy clusters at $z=0$, using a NVIDIA P100 GPU. We focused our tests on the version of the algorithm working on the identification of purely stellar substructures, \emph{stellar} {\tt SubDLe}.}
   {Our stellar {\tt SubDLe}  proved very efficient in identifying most of the galaxies, 82 per cent on average, in a set of 12 clusters at $z=0$. In order to prove the robustness of the method, we also performed some tests at $z=1$ and increasing the resolution of the input density grids. The average time taken by our {\tt SubDLe} to analyze one cluster is about 70 seconds, around a factor 30 less than the typical time taken by SubFind on a single computing node.}
   {Our stellar {\tt SubDLe} is capable of identifying the majority of galaxies in the challenging high-density environment of galaxy clusters in short computing times. This result has interesting implications in view of the possibility of integrating fast subhalo finders within simulation codes, that can take advantage of accelerators available on state-of-the-art computing nodes.}


   \titlerunning{ {\tt SubDLe}: Identification of substructures with deep learning}
   \authorrunning{Michela Esposito et al.}
   \maketitle
%

\section{Introduction}

\label{sec:introduction}
Cosmological hydrodynamical simulations are a fundamental tool to study the formation of cosmic structures. They solve the physics of gravity through N-body schemes while the more complex baryonic component is evolved through Eulerian or Lagrangian codes to describe hydrodynamics and sub-resolution models for additional processes happening on scales smaller than the typical resolutions, i.e. star formation and stellar feedback \citep[e.g.][]{springel03, muppi}, chemical enrichment \citep[e.g.][]{tornatore07,saitoh16}, radiative cooling \citep[e.g.][]{wiersma09}, growth of supermassive Black Holes and associated feedback \citep[e.g.][]{springel05,steinborn15} etc.

The fast progress in Computer Science has led to the development of a large number of reliable efficient codes for cosmological simulations. In order to fully exploit the predictive power of simulations, advanced analysis tools are required, especially structure finders.
In the already complex context of structure finding, a further complication has been added since substructures in the Dark Matter distribution have been found to survive within larger halos in simulations, at odds with what early simulations predicted due to numerical artifacts. This prompts the definition of two main classes of algorithms for structure finding: \emph{halo} finders, strictly speaking, are those which search for virialized structures around local maxima of the matter density field (the ``knots'' of the cosmic web); \emph{subhalo} finders are faced with the more challenging task to identify structures orbiting within a larger parent halo. Besides these, there are also algorithms designed to identify  filaments. 

Subhalo finding developed in a discipline itself during the past decades, given its crucial role of mapping the simulated phase-space distribution of matter to observable objects, like galaxies. This is a necessary step to study galaxy properties in simulations and compare them with observations. 

Though a proper classification of the algorithms for halo/subhalo finding cannot be easily defined, as reviewed in \citet{knebe13b}, they can be broadly divided in two categories stemming from the first two  well known algorithms developed for this purpose:
\begin{itemize}
    \item Direct collectors of groups of close particles, like the Friends-of-Friends (FoF) method \citep[][]{davis85}. Here the particles are grouped such that the interparticle distance is always below a parameter, the \emph{linking length}, usually given in units of mean interparticle distance. This is used as the first step of the strategy employed by the algorithm {\tt SubFind} \citep[]{subfind, stellarsubfind} to identify substructures; the FoF run allows to define the parent halos whithin which the actual search for substructures is performed. The linking length is usually assumed to be about 0.2 (the exact value being dependent on cosmology), since this was proved to identify halos with a density threshold similar to what is predicted for virialized systems by the spherical collapse model. The classical FoF algorithm, when applied to subhalo identification, has the disadvantage of linking together close structures through thin filaments of particles. Nevertheless, extension of the original FoF have been developed to account for this, for example ROCKSTAR \citep[]{rockstar}, which does a first separation of large groups in the classical FoF approach, then refines it by hierarchically searching for subgroups in 6D phase-space, while also tracking the temporal coordinate.
    \item Density peak locators complemented with a neighbor search, like the Spherical Overdensity (SO) method \citep[see][]{lacey94}. In this approach, the first step is the identification of ``locations'' of candidate halos, through a search of peaks in the density field or local minima in the potential field, among many possibilities. Then, the structures are grown around these locations by grouping neighboring particles, up to a given boundary, whose definition may vary a lot in different algorithms. The second step of the subhalo finding performed by {\tt SubFind} follows this approach, by looking for overdensities at the saddle points of the density field. Another example is SKID, described in details in \citet{skid}, where particles are repeatedly ``slid'' across the local gradient of the density field until they are collected around local maxima. 
    Then in both cases, the catalogue of subhalo candidates has to undergo an unbinding procedure, where particles that are not dynamically bound to their candidate subhalo are pruned from the catalogue. 

\end{itemize}
A detailed comparison of different halo and subhalo finders can be found in \citet{knebe11,knebe13a} and \citet{onions12} for Dark Matter only simulations and in \citet{knebe13b} for full-physics simulations, where the authors have investigated the source of scatter in halo properties among different codes. 
Halo and subhalo finders are now routinely built into the simulation codes, allowing for identification of subhalos on-the-fly, during the evolution of the simulation. This allows to save subhalo catalogues rather than the complete raw output of simulations and also access information on galaxy properties on-the-fly. 
Unfortunately, the overhead in computational cost when using such subhalo finders is usually non negligible, reaching up to the point where their cost over a significant number of snapshots becomes comparable to the simulation itself. This is due to the large number of particle-by-particle operations involved in the different steps of structure finding, like the search for saddle points in {\tt SubFind} or for the local maxima in SKID, which might also require several iterations. Furthermore, as the resolution of simulations increases, so do the number of particles and the computational cost of structure finders. Many algorithms have been updated or designed to work in parallel, like VELOCIraptor \citep{velociraptor}, which searches for substructures as peaks above the velocity distribution of the background, or {\tt SubFind} itself. However, the cost of running these algorithms while the simulation evolves is still a large limitation. On the other hand, performing frequent on-the-fly identifications of subhalos would be desirable for different applications. For example, one such applications could be to implement models for sub-resolution physics which rely on global properties of each single galaxy or Dark Matter subhalo. 

In this paper, we propose {\tt SubDLe} ({\tt Sub}structure identification with {\tt D}eep {\tt Le}arning),  an algorithm for substructure identification which is based on an application of methods of deep learning \citep[e.g., Chapter 3 of][]{NN}, a subclass of Machine Learning \citep[see][for an overview of Machine Learning techniques applied to astronomy]{ml}. Based on the extreme efficiency of deep learning methods in terms of computational resources, with this approach, we aim at drastically reducing the computational cost of identifying substructures in large cosmological simulations.

The application of the so-called \emph{supervised} techniques requires access to a ``training set'', consisting in input-output pairs, which define the problem to solve and the ideal solution. Then, through the optimization of an objective function, a set of parameters, which define how the algorithm solves a specific task, are iteratively adjusted in order to reproduce the solutions. This is the ``training'' stage which is the most computationally expensive step in the definition of the Machine Learning model, but only needs to be performed once if the data domain does not change significantly. A proper training requires the final model to be able to work outside the training set, i.e. it needs to be able to solve problems belonging to the same class of the training examples, but characterized by different data sets.

Artificial Neural Networks \citep[e.g., Chapter 1 of][]{NN} are a class of algorithms capable to approximate strongly nonlinear functions, with a flexible structure which makes them suitable for various tasks. Convolutional Neural Networks \citep[CNNs, e.g. Chapter 8 of][]{NN} are a network architecture tailored for visual tasks. For example, \citet{teodoro23} applied a CNN to clean galaxy catalogues of contaminants by identifying star formation regions or shreds of galaxies in large photometric surveys containing millions of sources.
Indeed, object detection with CNNs has been widely implemented in different fields and can be a suitable option to perform frequent on-the-fly identification of galaxies, given the computational advantages of a trained network.  

In this work we present an application of a specific CNN, {\tt U-Net} \citep[][]{unet}, to 3D density maps of galaxy clusters obtained from a set of cosmological hydrodynamical simulations carried out with the {\tt Gadget-3} code, an updated version of the cosmological hydrodynamical code {\tt Gadget-2} \citep[]{gadget2}. In our analysis we perform the training on subhalo catalogues provided by the built-in subhalo finder {\tt SubFind}. 
Our purpose is to  present our algorithm {\tt SubDLe}, an application of the {\tt U-Net}, as a subhalo finder, though it could be successfully trained to work in different instances of structure finders. 

The structure of the paper is as follows. In Section \ref{sec:methods} we describe the {\tt DIANOGA} set of simulations we used as training and test sets, the basics of CNNs and of the {\tt U-Net}, with details about the adopted training strategy for  {\tt SubDLe}. In Section \ref{sec:results} we report first the results of the trained model, for substructures that include DM, gas and stellar particles. We then focus on the case in which substructures are identified at redshift $z=0$ only from the distribution of star particles. Some tests are performed at $z=1$ and incrementing the resolution with respect to the fiducial one of the density grids. Conclusions and future perspectives of this work are summarized in Section \ref{sec:conclusions}.
    

\section{Methods}
\label{sec:methods}
\subsection{Simulations}
\label{sec:simulations}
The analysis presented in this paper is based on a set of 13 simulated galaxy clusters extracted from the {\tt DIANOGA} set of cosmological hydrodynamical simulations \citep[e.g.][]{bonafede11,rasia.etal.2015}. Initial conditions for this set of simulations consist of 29 Lagrangian regions centered on massive halos originally identified in a $1024^3$ Dark Matter (DM) particles box of 1$\,h^{-1}$Gpc comoving side. These initial conditions have been generated with the \emph{zoomed-in initial conditions} technique \citep[][]{tormen97}. The adopted cosmology is a flat $ \Lambda CDM$ with: $h=0.72$, $\Omega_m=0.24$, $\Omega_b=0.04$, $n_s=0.96$ and $\sigma_8 = 0.8$. These regions were re-simulated with a developer version of the TreePM-SPH code {\tt Gadget-3} \citep[evolution of {\tt Gadget-2},][]{gadget2}. 
The mass resolution for DM and gas particles is $m_{DM}= 8.47 \times 10^8 h^{-1} \text{M}_{\odot}$ and  $m_{gas}= 1.53 \times 10^8 h^{-1} \text{M}_{\odot}$, respectively, while the Plummer-equivalent softening lengths are $\epsilon= 5.6 \ h^{-1}$ kpc for DM and gas, and $\epsilon_{\ast}= 3 \ h^{-1}$ kpc for stars.
The simulations analysed here include the description of the following processes: metal-dependent radiative cooling, star formation, chemical enrichment from stellar evolution and feedback from supernovae and Active Galactic Nuclei. We refer to  \citet{ragonefigueroa13} for a detailed description of this set of simulations.

In the simulated regions, galaxies have been identified with the {\tt SubFind} algorithm \citep{subfind,stellarsubfind}. As already mentioned in Section \ref{sec:introduction}, this subhalo finder starts to look for substructures with a FoF search, that identifies DM halos. Within them, substructures are then identified by looking for regions enclosed by density contours crossing a saddle point of the density field. After that, particles belonging to each of the subhalo candidates undergo an unbinding procedure to exclude those particles that are not gravitationally bound to the assigned subhalo. In Section \ref{sec:results} we will refer to the clusters we have analysed as {\tt CL-$i$}, where $i$ runs from 1 to 13. The main properties of these clusters are reported in Table \ref{tab:clusters}.
\begin{table}
	\centering
	\caption{Main properties of the clusters analysed in this study, including their virial mass, virial radius and the phase of the analysis in which they were used (i.e. training, referred to as ``train'', validation, referred to as ``val'', or test, see below for details)}
	\label{tab:clusters}
	\begin{tabular}{lcccr} 
		\hline
            \\
		ID & $M_{vir}$ & $R_{vir}$ & $N_{gal}$ & phase\\ 
         & [$10^{15} \ \text{M}_{\odot}$] & [Mpc]& & \\
		\hline
            \\
            {\tt CL-1} & 0.74 & 2.4& 192 &train/  val\\
            {\tt CL-2} & 0.73 & 2.4& 231 & test\\
		{\tt CL-3} & 2.3 & 3.4& 571 &test\\
		{\tt CL-4} & 0.92 & 2.5& 236&test\\
		{\tt CL-5} & 0.24 & 1.6& 75 &test\\
		{\tt CL-6} & 2.1 & 3.3& 510 &test\\
      	{\tt CL-7} & 0.18 & 1.5& 40 &test\\
        {\tt CL-8} & 1.7 & 3.1& 505 &test\\
        {\tt CL-9} & 1.9 & 3.2& 604 &test\\	
        {\tt CL-10} & 2.4 & 3.5& 602 &test\\		
        {\tt CL-11} & 2.3 & 3.5& 535 &test\\
        {\tt CL-12} &  2.2 & 3.4& 543 &test\\
        {\tt CL-13} & 2.3 & 3.2 & 542 &test\\
        \\
		\hline
	\end{tabular}
\end{table}

\subsection{Convolutional Neural Networks}
\label{cnn}
Convolutional Neural Networks (CNNs) are a class of Machine Learning algorithms with a multilayer architecture \cite[][Chapter 8]{NN}, where the (local) input is mapped into the (local) output through a series of linear or nonlinear operations, in each computational layer. 
CNNs are designed to work with grid-like inputs, like images, text, time series or data sequences, which exhibit strong variations.
 Image data are translation invariant as far as object identification is concerned, since any object in a given image has the same interpretation independently of its position within the image itself. This makes images the ideal input for CNNs, since they are inherently translation invariant.  This can be explained as the basic steps performed by a CNN operate on local regions of the input grid and depend only on relative spatial coordinates. This is what we need for substructure identification, since we want to be able to identify galaxies anywhere in the simulated region.

CNNs are defined as Artificial Neural Networks that use a \emph{convolution} operation (see Appendix \ref{sec:appendix}) in at least one layer. 
A network with only layers involving convolution operations is called a Fully Convolutional Network \citep[FCN,][]{fcn}, which is the case of the {\tt U-Net} chosen for this work.

 In this Section we describe the {\tt U-Net}, which we have applied in a generalized 3D version to approach the problem of identification of substructures in cosmological simulations (Section \ref{sec:u-net}), in order to provide the basic knowledge of how the subhalo finder works and where its advantages and limitations come from (see Section \ref{sec:results}).

\subsection{{\tt U-Net}, data set and training}
\label{sec:u-net}
{\tt U-Net} is a FCN developed to perform segmentation in biomedical images. Segmentation is a task involving classification of each pixel of an image in order to define the contours of objects and not only a box around the target, as in classical object detection tasks. Fig. 1\footnote{\url{https://lmb.informatik.uni-freiburg.de/people/ronneber/u-net/}} of \citet{unet} shows the architecture of the network which includes the basic ingredients of a CNN, described in Appendix \ref{sec:appendix}. We refer to \citet{unet} for a detailed description of the philosophy of its architecture and the training strategy. For our analysis, we modified the original architecture of the {\tt U-Net} in order to work on the identification of subhalos in 3D, in cosmological hydrodynamical simulations. In our application, the pixel values represent the mass densities.
We built a 3D version of {\tt U-Net} using the pyTorch\footnote{\url{https://pytorch.org/}} library. 

\begin{figure}
    \centering
    \includegraphics[width=0.5\textwidth]{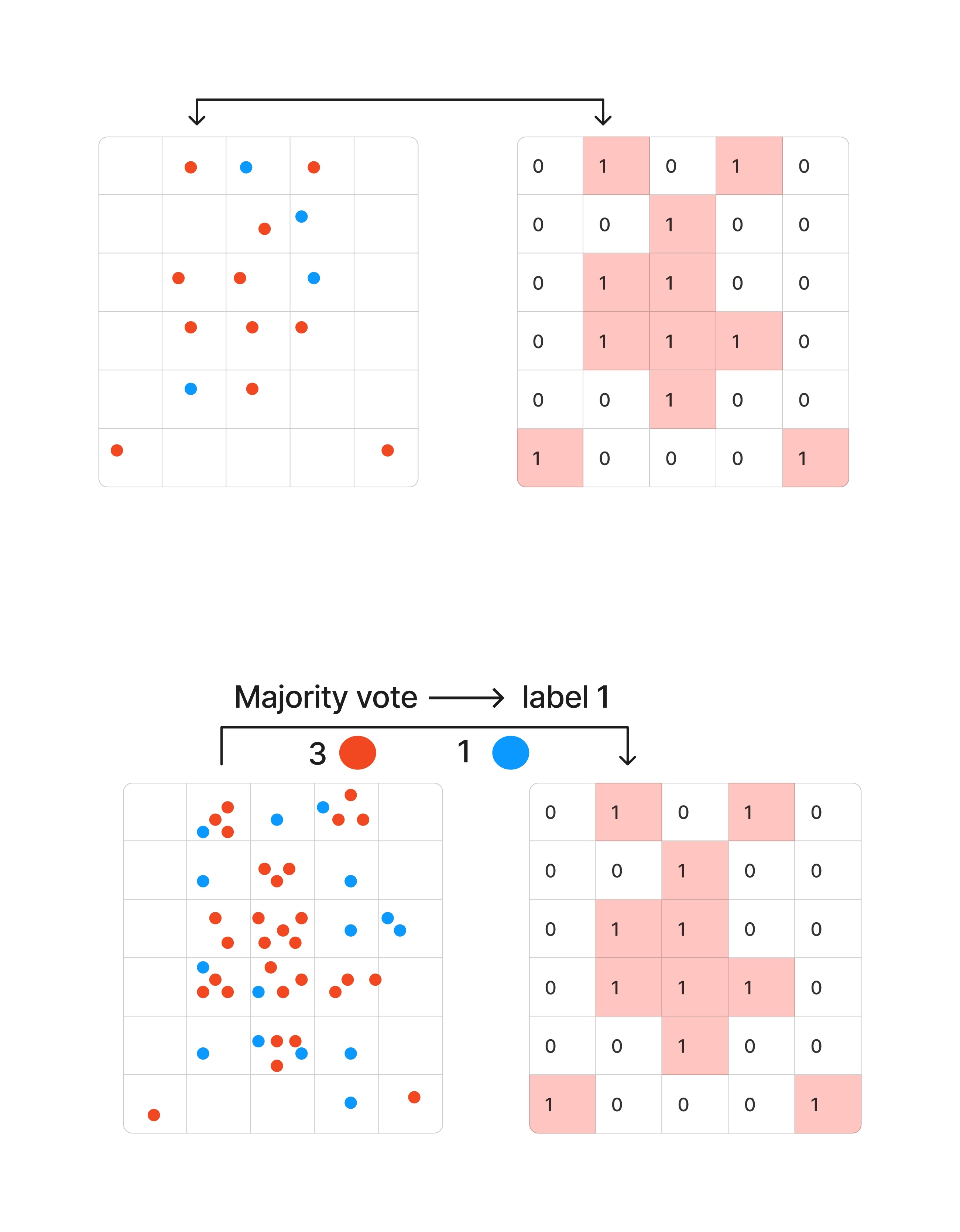}
    \caption{Example of the assignment scheme of labels adopted in the training set of the 3D {\tt U-Net} within {\tt SubDLe}. Here red particles trace the substructures to be identified, and blue particles are part of a background. In the upper panel, at most one particle is present in each grid pixel, thus the substructures (associated with the label ``1'' in the right-hand grid, whereas the background is labeled as ``0'') are simply those containing one red particle. In the lower panel, more than one particle is allowed in each grid pixel, so the assignment of labels is performed with a majority vote.}
    \label{fig:assignment} 
\end{figure}

The training of the  3D {\tt U-Net} that performs most of the calculations in {\tt SubDLe}  requires a set of reliable data to ``teach'' the network how to identify substructures. This is done by using the subhalo catalogues produced by {\tt SubFind} in the simulations described in Section \ref{sec:simulations}. 
For our aims, using the direct {\tt SubFind} output, which associates each particle to a substructure (or to none, if the particle is not dynamically bound to any subhalo) is not feasible. The reason is that a CNN works with grid-like inputs with a fixed dimension; the particles of a halo instead define a sparse data structure. We thus mapped our {\tt SubFind} output to grids having fixed dimensions. A grid is defined as a box having an origin in a given point of space ($X_{g}$, $Y_{g}$, $Z_{g}$), a size $L_{g}$, and $N_{g}$ points per dimension.
If a particle has coordinates in the ranges $[X_{g}, \ X_{g}+L_{g}]$, $[Y_{g}, \ Y_{g}+L_{g}]$, $[Z_{g}, \ Z_{g}+L_{g}]$, it belongs to the grid and will fall in a grid pixel.
Let us consider the simple situation in which each grid pixel contains at most one particle, as in the upper panel of Fig. \ref{fig:assignment}. The red particles as those associated, according to previous knowledge (i.e. {\tt SubFind}'s identification), to the objects we want to identify, whereas the blue particles are just part of a ``background''. Then, we can assign the label ``1'' to the substructures and ``0'' to the background. Thus, in Fig. \ref{fig:assignment}, grid pixels which contain red particles are assigned the label ``1'' in the right grid of the top panel, containing labels for all grid pixels, while all the others are assigned the label ``0''. Typically, more particles will fall in the same pixel; then, the pixel will be assigned to a substructure or to the background using a majority vote, as in the lower panel of Fig. \ref{fig:assignment}. In the unlikely case of parity, we are still assigning the pixel to a substructure. We refer to the grid carrying labels as ``s-grid''. 

For the problem of identification of substructures in simulations, a twin grid simply containing the stellar mass density in the grid region is also computed, and we call it ``d-grid'', which we have built with a Nearest Grid Point (NGP) assignment scheme, for simplicity. This is a case of ``binary segmentation'', in which a grid pixel is classified as a member of one class (label ``1'', denoting substructures in our case)  or not (label ``0''), resulting in s-grids containing only 1s and 0s, as in the simplified examples shown in Fig. \ref{fig:assignment}. 

The training of the {\tt U-Net}, where the parameters of the filters (see Appendix \ref{sec:appendix})  are iteratively adjusted so as to maximize the similarity of the predictions of the model to the solutions extracted from {\tt SubFind}'s catalogues,  was performed with one NVidia P100 GPU with 16 GB memory, provided by the online data science platform Kaggle\footnote{\url{https://www.kaggle.com/}}. The training, as always for Neural Networks \citep[][ Section 1.2.1]{NN} is done through the minimization of a ``loss function '', which defines the error of the model's prediction at each iteration. In simple cases of regression the chosen loss function is often a simple Mean Squared Error. The minimization is performed with respect to the free parameters of the filters, with backpropagation \citep[see][Section 1.3]{NN}. Since it is conceptually wrong to evaluate the best set of parameters on the same data set on which the optimization of the loss function is performed, a set of data is reserved for validation, over which the loss function is evaluated and the best model is selected from the set of parameters that give the minimum loss function on the validation set.  Our chosen loss function is the Binary Cross Entropy, a version of the Cross Entropy loss function, adopted in \citet{unet}, for cases of binary segmentation, such as ours. 

Due to memory constraints at the training stage, we needed to cover each of our simulated clusters with many sub-grids. In fact, a single grid covering the whole cluster would require too much RAM for achieving the minimal spatial resolution required to resolve the smallest substructures with a few pixels. Clearly, the dimension of each grid, with which a cluster region is sampled, also determines the largest scale on which the identification can benefit from the network's architecture, since identifying structures larger than $L_{g}$ is problematic. In this case, indeed, the large structures will be split into more than one grid, thereby forcing our {\tt SubDLe} to try to identify ``shreds'' of a substructure. However, we emphasize that this represents only a technical limitation, strictly determined by the accessible hardware, and not an intrinsic limitation of the method. 

We adopted a number of grid meshes per side $N_{g}$=64, following the convention for images of having a power of 2 as the number of pixels per side 
and the physical dimension of pixels was chosen to be of about 10 kpc in order for a full d-grid to be larger than average galaxies. Unless otherwise stated, in Section \ref{sec:results}, we always assume a resolution of about 10 kpc, with small variations (ranging from 9 to 12 kpc), which make little to no difference in the final subhalo catalogue. 

The {\tt U-Net} is trained using both s- and d-grids, by learning how to assign d-grid pixels to substructures as defined by the s-grid. 

 The {\tt U-Net}  part of {\tt SubDLe}  is designed to output a $64^3$ grid with pixel values defining the probability $p$ of each pixel to belong to a substructure (the class labeled as ``1''). A threshold is then chosen in order to decide which pixels are assigned to substructures. The most common choice is a threshold of $p_{th}=0.5$, so we adopted it without tuning for this parameter. Then all pixels with a probability $p>0.5$ are assigned the label ``1'', whereas the remaining ones are labeled as background (label ``0''). This outputs an s-grid containing recovered substructures (at this stage, these are actually a list of pixels assigned the label ``1''). 

\begin{figure*}
    \centering    \includegraphics[width=\textwidth]{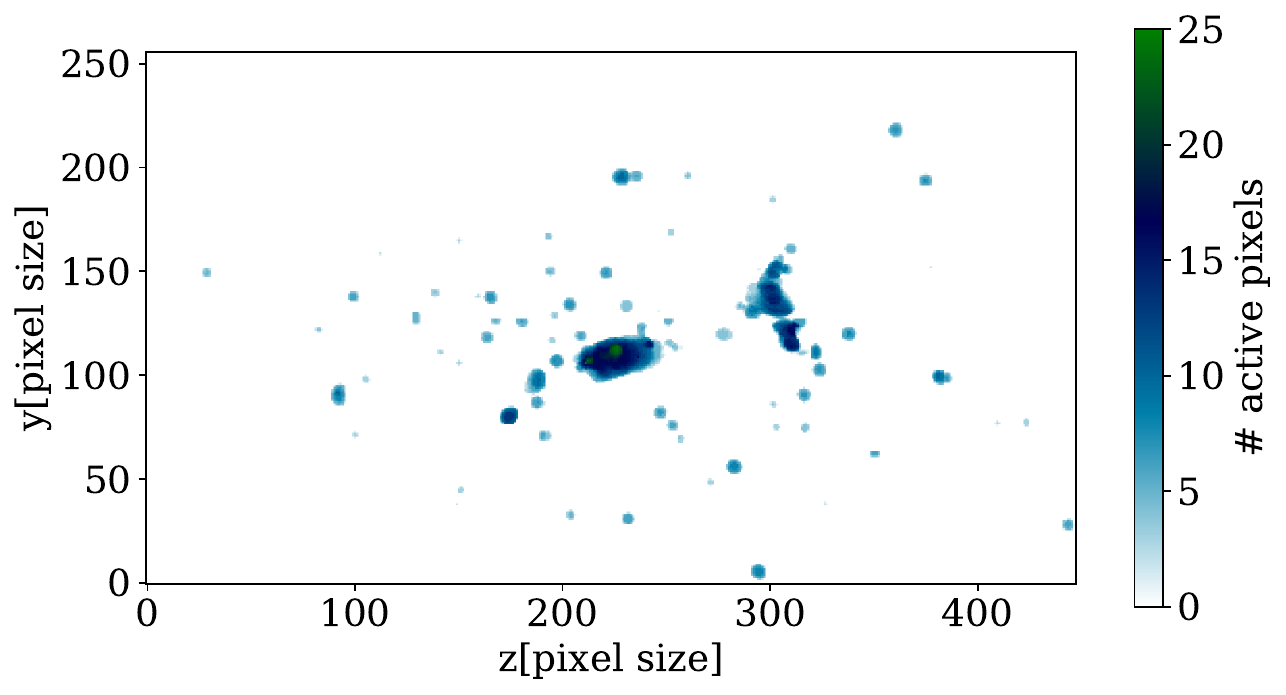}
    \caption{Projected map of the active pixels associated to the substructures identified by  the first step of the identification performed by {\tt SubDLe} (through a 3D {\tt U-Net}) in  a run on a whole cluster, {\tt CL-5} at $z=0$. The blobs of active pixels (non-white, see the colormap on the right) need to be grouped together to identify a given subhalo, since at this stage all subhalos are associated to the same class label.}
    \label{fig:fof} 
\end{figure*}

Once pixels with a high probability to belong to a substructure are selected, as a final step we need to re-group pixels of all s-grids which can be attributed to the same subhalo (since \emph{all}  the subhalos are assigned the same label in the s-grids).
Fig. \ref{fig:fof} shows an example of output of {\tt SubDLe}  at this stage of the identification process. The map shows the 2D projection of density of active pixels (i.e. pixels with label ``1''), reconstructed from the s-grids covering the whole cluster {\tt CL-5} at $z=0$ (see Section \ref{sec:simulations}).

 {\tt SubDLe} performs  the grouping of  active  pixels in the s-grid relying on a proximity criterion through a Friends-of-Friends algorithm implemented with pyfof \footnote{\url{https://pypi.org/project/pyfof/}}. The linking length was fixed to $\sqrt{3}$ times the pixel size, in order to group contigous pixels along all directions. After this stage we have a substructure catalogue defined on our grid. At last, original particles are assigned to recovered substructures by inverting the NGP grid assignment.


\section{Results}
\label{sec:results}

\subsection{  {\tt SubDLe} }

\label{sec:results0}
In this Section we present the results from the application of {\tt SubDLe}  to the total mass density 3D maps of galaxy clusters, including the contribution of DM, gas and stellar particles, and compare the output subhalo catalogue with that based on the application of {\tt SubFind}.

The training of the {\tt U-Net}  within {\tt SubDLe}  is performed on approximately 1000 d-grids (700 dedicated to the training itself and 300 to the validation) extracted from the the cluster {\tt CL-$1$} at $z=0$, with a total virial mass of $\text{M}_{\text{vir}} \approx 7.4 \times 10^{14} \ \text{M}_{\odot}$. Each $64^3$ d-grid extracted from it was visually inspected in order to provide highly representative training data.  Even though we expect an improvement in the performance when using a larger training data set, we do not deem it necessary for the purposes of the exploratory study presented in this work. In any case, we perform some preliminary tests based on a larger training data set, as discussed in the following, in Section \ref{sec:2ndtraining}. 

\begin{figure*}
 \centering
 \begin{subfigure}[b]{0.4\textwidth}
   \includegraphics[width=\textwidth]{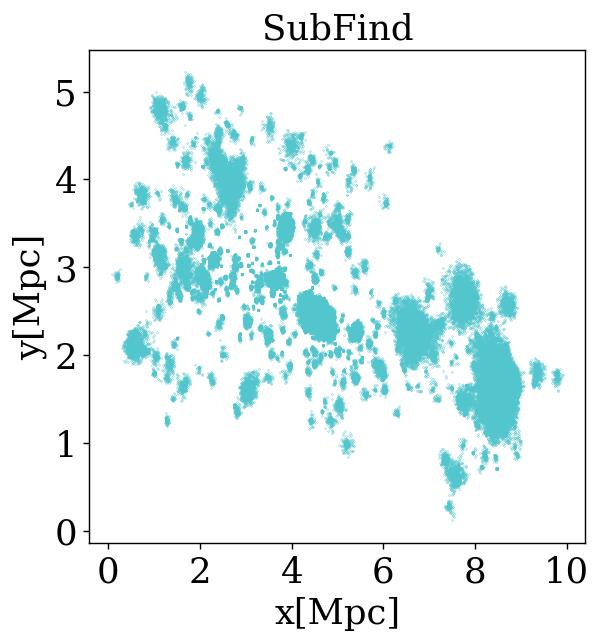}   
 \end{subfigure}
 \begin{subfigure}[b]{0.4\textwidth}
  \includegraphics[width=\textwidth]{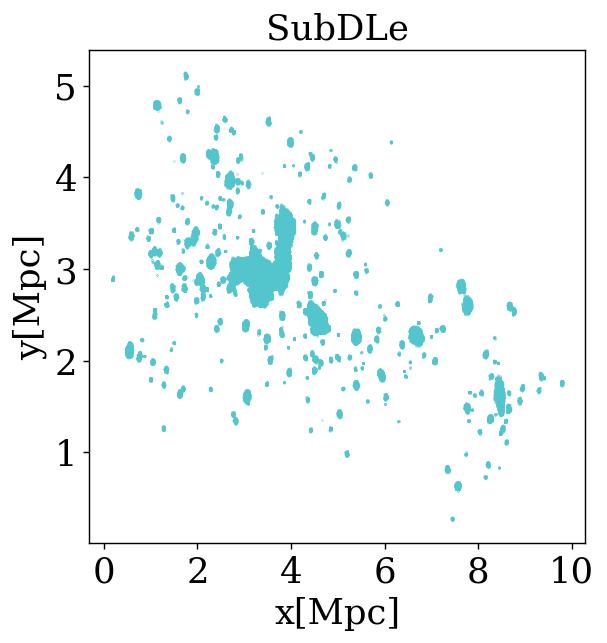}

 \end{subfigure}
 
 \caption{\label{fig:D4}Projected map of  particles belonging to subhalos of the cluster {\tt CL-2} at $z=0$.  The left and right panels show the subhalos identified by {\tt SubFind} and   {\tt SubDLe}, respectively.}
 
\end{figure*}

The trained {\tt SubDLe} was tested on the galaxy cluster {\tt CL-2} at $z=0$, with approximately the same virial mass as {\tt CL-1} (see Table \ref{tab:clusters}).  Here the d-grids are extracted so that they overlap with their neighboring grids by half their sides in all three directions in order to properly sample border regions, for a total of 6851 grids covering a box of about $(10 \times 5 \times 5 ) \ {\rm Mpc}^3$. Each d-grid covers a cubic region of about 640 proper kpc per side. Here the mass density sampled by the d-grids includes \emph{all} the particles contained in {\tt SubFind}'s catalogues, i.e. DM, stars and gas particles. A visual inspection of the results can be done by comparing the two panels of Fig. \ref{fig:D4}.

The left panel of Fig. \ref{fig:D4} shows the projected distribution of particles belonging to different substructures identified by {\tt SubFind} in the cluster {\tt CL-2}, whereas the right panel shows the substructures identified by {\tt SubDLe} in the same cluster. Qualitatively, {\tt SubDLe} appears to be able to locate most of the major substructures, while missing some of the smaller ones. The latter are inevitably described by a weaker signal in the s-grids. On the other hand, the spatial extension of   {\tt SubDLe} 's substructures is generally smaller than the size of corresponding {\tt SubFind} substructures. This is due to the fact that the {\tt SubDLe} more easily identifies high density constrast regions. The centers of subhalos have the maximum density contrast, and the network tends to miss external low-density cells, even if dynamically they are bound to the same subhalo. Nevertheless, the hardest task in substructure identification is the location of density peaks, which {\tt SubDLe} performs pretty well, as we will discuss more quantitatively later in this Section, whereas the detailed matter distribution of each subhalo can be recovered later with a fast neighbor search algorithm, as mentioned in Section \ref{sec:introduction} for the SO method. Another detail that emerges from the comparison of the distribution of {\tt SubDLe} and {\tt SubFind} substructures in Fig. \ref{fig:D4} is the presence of a very large subhalo among {\tt SubDLe} sustructures at $(x,y) \approx (3 \ {\rm Mpc},3 \ {\rm Mpc})$, while the corresponding region in the left panel of Fig. \ref{fig:D4} is devoid of large structures. 
This corresponds to the largest central subhalo of the cluster, hosting the Brightest Cluster Galaxy (BCG).  This does not appear in {\tt SubFind}'s substructures because this algorithm mixes the largest subhalo of a FoF group with the ``fuzz'' of particles which are not linked to any specific subhalo in the first step of the identification process \cite[see][]{subfind}. {\tt SubDLe} can thus also be applied to overcome this limitation of {\tt SubFind}, though further specific tests are required to assess this application. A Machine Learning algorithm developed to disentangle the BCG from the IntraCluster Light in the distribution of star particles in simulations of galaxy clusters, based on their dynamical properties, has already been developed and tested in \citet{marini22}. In the following we will focus on   {\tt SubDLe}'s ability to reproduce the average subhalo population in cluster-sized halos. We did not perform tests in different environments, but we can only expect better results in less dense environments where galaxies are more isolated and thus more easily separated from the background by a visual tool like {\tt SubDLe}.

\begin{figure}
\centering
\includegraphics[width=0.475\textwidth]{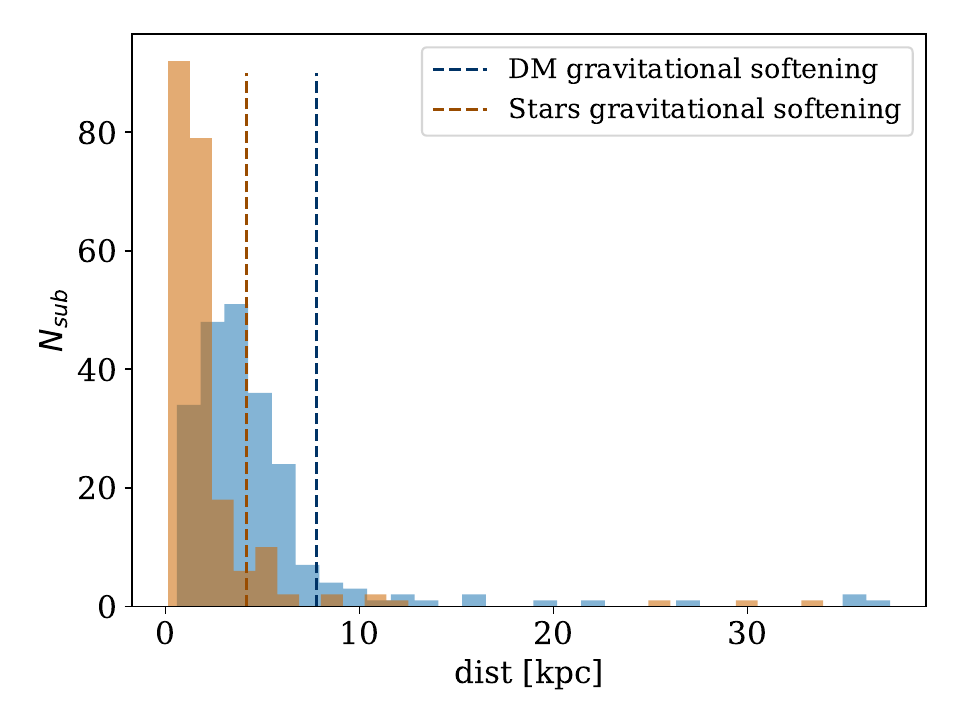}
\caption{\label{fig:histogramd4_dist_comp} Distribution of pairs of {\tt SubFind}/{\tt SubDLe}  subhalos as a function of the distances between their centers, in proper kpc, for the identification performed by the all-particle {\tt SubDLe}  (in blue) and stellar   {\tt SubDLe}, in orange. The vertical dashed lines mark the gravitational softening used in the gravitational force calculations for the DM particles (thus the majority of the particles, in blue) and for star particles (in orange).}
\end{figure}

In order to quantitatively compare the catalogues produced by {\tt SubFind} and  {\tt SubDLe}, we matched each subhalo found by {\tt SubDLe} to the closest one identified by {\tt SubFind}. Here only {\tt SubFind} subhalos  resolved with at least 100 particles are included. {\tt SubDLe} substructures are processed in decreasing order of total mass, so as to avoid matching a {\tt SubFind} subhalo to a close small clump of particles found by {\tt SubDLe} before the actual corresponding {\tt SubDLe} subhalo is processed. The proximity is estimated through the distance between the {\tt SubFind} center of a subhalo, identified as the position of the most bound particle in each substructure, and the {\tt SubDLe} center, which we define as the center of mass of the subhalo, for simplicity.
The search for matching substructures is performed within a radius of 50 kpc, then the matched {\tt SubDLe} substructures which do not share any particle with their closest match are excluded from the catalogue, since they are unlikely to be matching any {\tt SubFind} subhalo.
Since {\tt SubDLe} tends to miss the external lower-density region of subhalos, while detecting the central density peaks, as shown in Fig. \ref{fig:D4}, the criterion of requiring at least one common particle is enough to guarantee a significant overlap and thus a correct match between {\tt SubFind}'s and   {\tt SubDLe}'s subhalos. We are interested in   {\tt SubDLe}'s performance mainly as a peak locator, so we do not assume a more strict criterion.
The distribution of distances between the center of a {\tt SubDLe} substructure and the center of the corresponding {\tt SubFind} one is shown in Fig. \ref{fig:histogramd4_dist_comp}, in blue.

We find that 91 per cent of pairs of substructures have distances below the gravitational softening of the DM particles, roughly representing the spatial resolution of the simulation. There is a tail of pairs which expands up to few 10s of kpc, which might be due to a displacement between the center of mass and the center of the gravitational well or to an artificial merging of close substructures as we will see below. Overall, 80 per cent of all the resolved substructures (i.e. with more than 100 total particles) identified by {\tt SubFind} have been identified by {\tt SubDLe} as well.

\begin{figure*}
 \centering
 \begin{subfigure}[]{0.3\textwidth}
  \centering  \includegraphics[width=\textwidth]{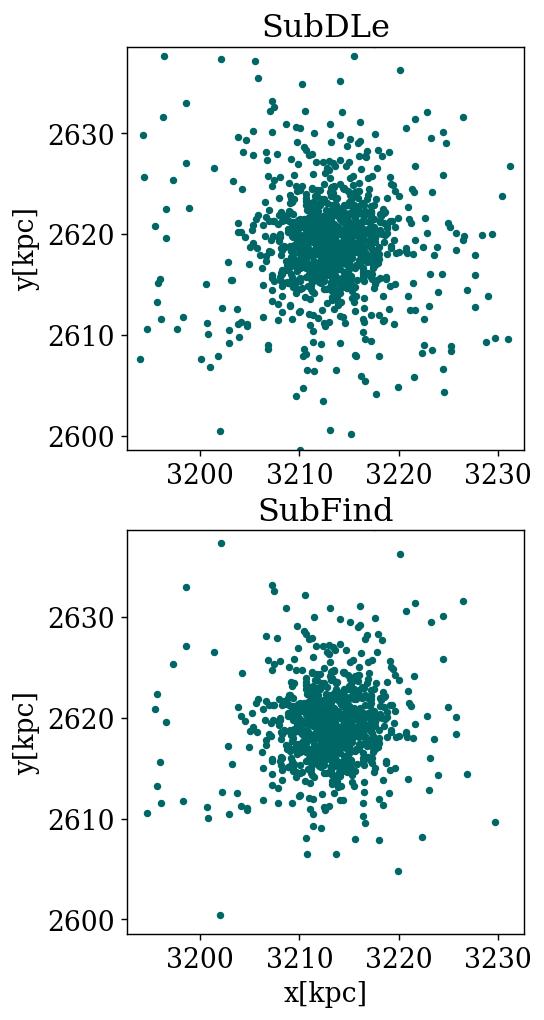}
  \label{fig:D4sub}
 \end{subfigure}
 \begin{subfigure}[]{0.3\textwidth}
   \centering   \includegraphics[width=\textwidth]{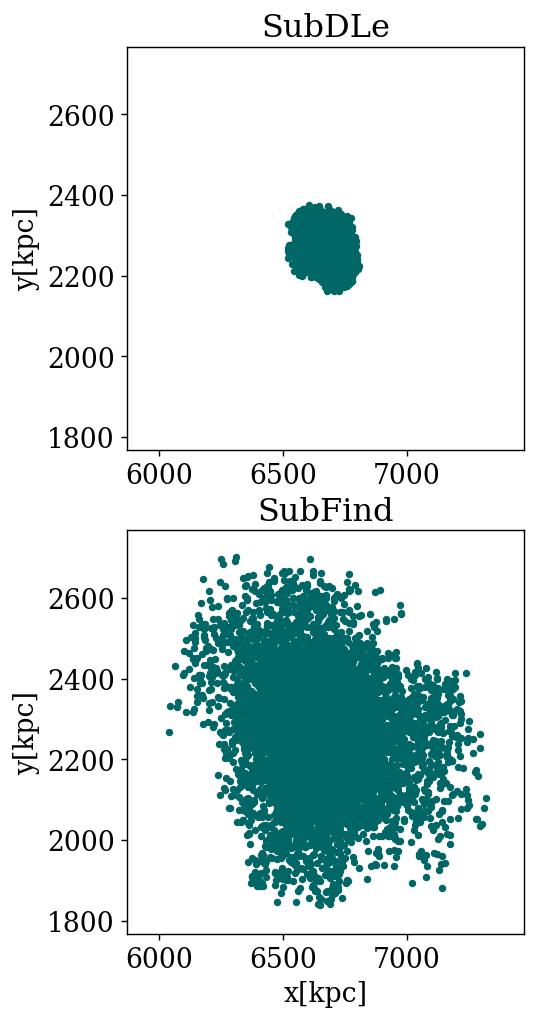}
   \label{fig:D4subcore}
 \end{subfigure}
  \begin{subfigure}[]{0.3\textwidth}
  \centering
  \includegraphics[width=\textwidth]{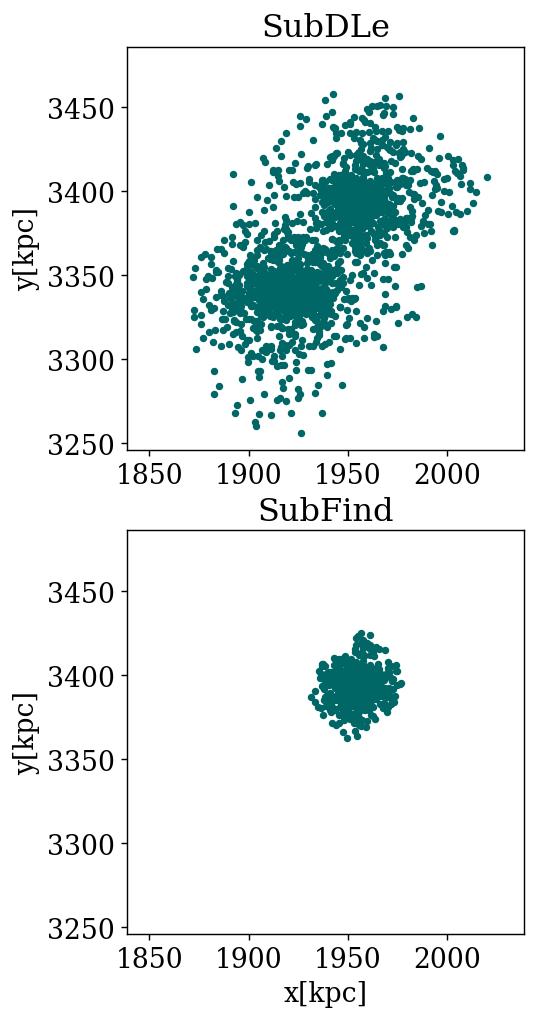}
  \label{fig:D4submerge}
 \end{subfigure}
 \caption{ \label{fig:d4cases} Projected spatial distribution of all particles (DM, gas and stars) belonging to a subhalo identified by {\tt SubDLe} (top row) and the matching subhalo identified by {\tt SubFind} (bottom row) for three typical cases: in the left column, a subhalo with an extension of few tens of kpc, perfectly identified by   {\tt SubDLe}; in the central column a larger subhalo, spanning a distance of about 1 Mpc, where only the central nucleus is identified by {\tt SubDLe}; in the right column, a system of two close substructures that are combined into one subhalo in   {\tt SubDLe}'s catalogue due to their short distance in comparison with the density grid resolution.}
\end{figure*}

Fig. \ref{fig:d4cases} shows the comparison between matched pairs of {\tt SubDLe}/{\tt SubFind} substructures in three representative cases:

\begin{itemize}
 \item The left column shows an ideal case, with a close to perfect match between the identifications performed by {\tt SubFind} and {\tt SubDLe} (98 per cent of the mass of the original subhalo is recovered by {\tt SubDLe}). Bottom and top panels show {\tt SubFind} and {\tt SubDLe} subhalos, respectively. The size of the subhalo, which is few tens of kpc, is well reproduced, while single particles have been added or missed by   {\tt SubDLe}, due to the lack of direct dynamical information on them in the input d-grids. Note that this subhalo is well reproduced despite being resolved with few pixels per dimension.
 \item The central column shows the already mentioned trend of {\tt SubDLe} cropping the most extended subhalos, while leaving only the higher density core. Note that this substructure extends for a few hundreds of kpc, thus it was probably split in multiple grids, reducing its detectability as a whole structure. In principle, we consider this limitation as not representing a major obstacle in the application of {\tt SubDLe}, since the most challenging task of locating the peak of the density field has been accomplished also in this instance. 
 \item The right column shows a case in which two visibly (and dynamically, as {\tt SubFind} identifies them as two different subhalos) distinct substructures are merged together by {\tt SubDLe} because of their proximity. Their centers are indeed few resolution elements away and {\tt SubDLe} is not able to disentangle them on the basis of their mass distribution alone. This is an effect of the limitation on the computational resources we accessed, since we cannot increase the resolution while keeping fixed the extension of d-grids for all clusters and fit into the GPU memory during the training. This problem would be attenuated if we repeated the training with more than one GPU available. The artificial merging of close substructures is one of the culprits of missing subhalos in   {\tt SubDLe}'s catalogues. We will discuss later in this Section how the increase of resolution can indeed boost the percentage of matched subhalos, by separating close, but distinct substructures.
\end{itemize}

\subsection{Stellar {\tt SubDLe} }
\begin{figure*}
 \centering
 \begin{subfigure}[b]{0.4\textwidth}
   \centering
   \includegraphics[width=\textwidth]{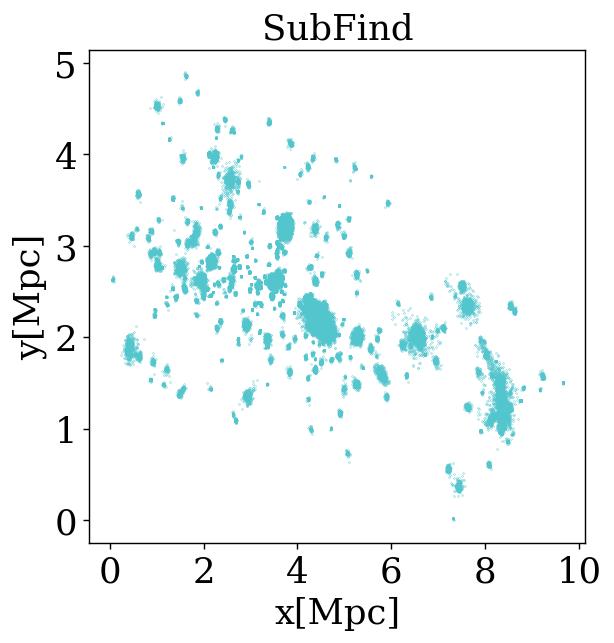}
   \label{fig:D4truestars}
 \end{subfigure}
 \begin{subfigure}[b]{0.4\textwidth}
  \centering
  \includegraphics[width=\textwidth]{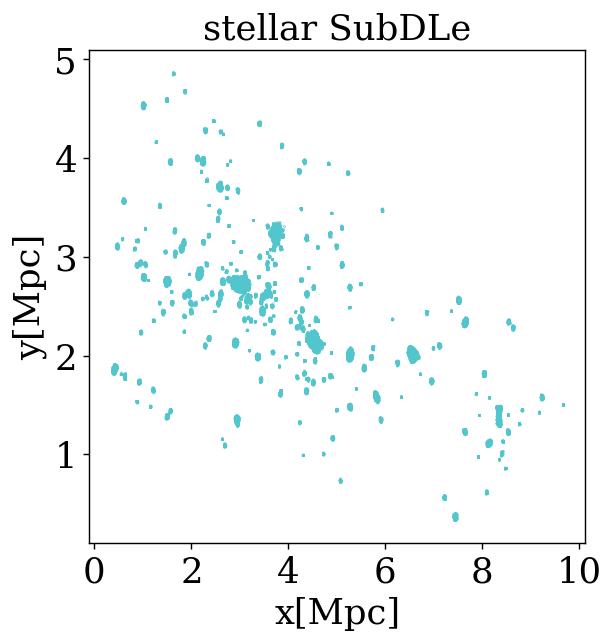}
  \label{fig:D4cnnstars}
 \end{subfigure}
 \caption{Projected distribution of star particles belonging to different galaxies  identified by {\tt SubFind} (left panel) and by {\tt SubDLe} (right panel) in the cluster {\tt CL-2} at $z=0$.}
 \label{fig:D4stars}
\end{figure*}

Since our network works inherently better in identifying highly contrasted clustered structures, we repeated the identification of substructures using stellar density maps instead of total density maps; we re-trained {\tt SubDLe} to do so and will refer to this version as \emph{stellar} {\tt SubDLe}. Stellar substructures, i.e. galaxies, are indeed more clustered in comparison to DM and gas structures due to their formation through dissipative collapse of gas clouds. The training data set consists of 700 grids of stellar mass densities extracted from cluster {\tt CL-1}, while 300 grids are used for validation. Fig. \ref{fig:D4stars} shows the same comparison of Fig. \ref{fig:D4}, this time with galaxies identified by {\tt SubFind} (left panel) and by the stellar {\tt SubDLe} (right panel) in the cluster {\tt CL-2} at $z=0$. The discrepancy in substructure sizes is still visually noticeable, but less severe than in the former case, where all particles were considered. The matching between {\tt SubFind} and stellar {\tt SubDLe} substructures is performed as described in Section \ref{sec:results0} and the distribution of distances is shown in Fig. \ref{fig:histogramd4_dist_comp}, in orange, and compared to the distribution for the {\tt SubDLe} run on all particles, in blue. The dashed lines define the softening lengths for star (orange) and DM (blue) particles. As expected, the centering is slightly more accurate in terms of absolute distances, but overall the percentage of galaxies for which the distances between matched pairs is below the gravitational softening for star particles is still 91 per cent, as for the run on all particles, with respect to the DM softening.

In total, the percentage of {\tt SubFind} galaxies resolved with at least 100 star particles (corresponding to a total stellar mass of about $4 \cdot 10^9 \  \text{M}_{\odot}$), identified by our stellar {\tt SubDLe}, is 93 per cent, thus sensibly larger than in the previous case. Besides identifying a larger fraction of substructures, the stellar {\tt SubDLe}  performs better in reconstructing the stellar mass distribution in comparison to the version based on all the particle species, as shown in Fig. \ref{fig:D4stars}. 
In order to quantify this, we computed the ratio between the mass in star particles which both {\tt SubFind} and{\tt SubDLe} assign to a matched pair of galaxies, $\text{M}_{\tt SF/SD}$, and the ``true'' mass. We assume the true mass to be the one given by {\tt SubFind} catalogues ($\text{M}_{\tt SF}$) for the purpose of this comparison. With the ratio $\text{M}_{\tt SF/SD} / \text{M}_{\tt SF}$  we quantify how complete the identification of a subhalo is in comparison with {\tt SubFind}, in terms of its mass distribution, but not how pure it is. The mass of {\tt SubDLe}'s subhalo is generally smaller than the true mass, because, as already mentioned, the  model  tends to identify only the cores of subhalos. Nevertheless, in cases similar to the one shown in the right column of Fig. \ref{fig:d4cases} which might still occasionally occur in the identification of stellar substructures, the mass of the {\tt SubDLe} ``galaxy'' is the sum of the two cores merged together. This is caused by the resolution of the grids and does not represent a conceptual flaw of our method. On the other hand, we are interested in quantifying the difference of subhalo masses caused by the intrinsic nature of the   {\tt SubDLe}, which does not ``see'' low-density regions in the d-grids. For this reason,   we remove the ``over-merging'' effect in the analysis of the reconstructed mass distribution, by considering only the mass of common particles in {\tt SubFind}/ {\tt SubDLe}  pairs ($\text{M}_{\tt SF/SD}$), rather than the total mass of each (possibly composite) substructure found by  {\tt SubDLe}. In other words, we are not interested in the ``purity'' of the mass distribution of {\tt SubDLe} galaxies in this step of the analysis. 

\begin{figure}
\centering
\includegraphics[width=0.475\textwidth]{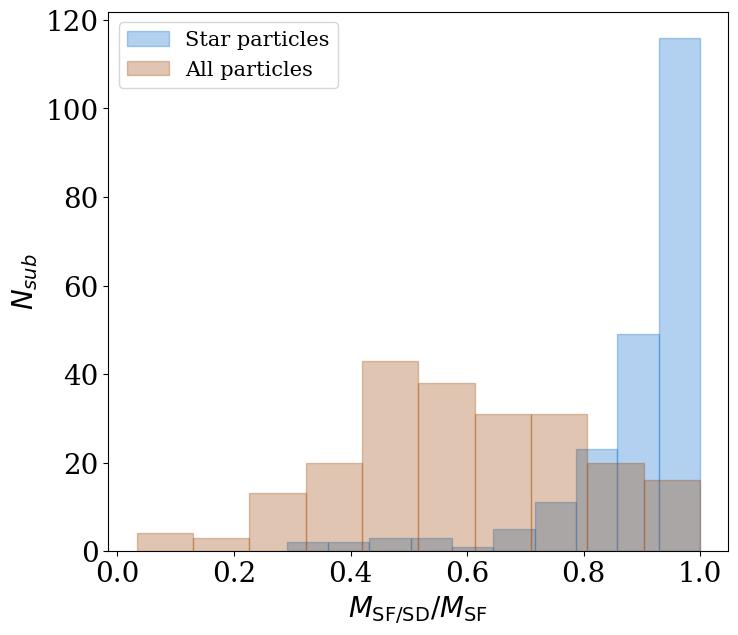}
\caption{\label{fig:histogramd4} Distribution of pairs of {\tt SubFind}/{\tt SubDLe}  subhalos as a function of the ratio between the mass of common particles ($\text{M}_{\tt SF/SD}$) and the mass given by {\tt SubFind} ($\text{M}_{\tt SF}$), as a tracer of the completeness of the mass reconstruction performed by   {\tt SubDLe}, for the run carried on the mass distribution of all particles (DM, gas and stars), in orange, and for the run of stellar {\tt SubDLe}, in blue, for the cluster {\tt CL-2} at $z=0$.}
\end{figure}

The distribution of $\text{M}_{\tt SF/SD}/\text{M}_{\tt SF}$  is shown in Fig. \ref{fig:histogramd4}, for matches of {\tt SubFind}/{\tt SubDLe}  substructures in {CL-2} when {\tt SubDLe} is applied to all particles (yellow) and on star particles only (blue), in both cases considering matching {\tt SubFind} substructures with at least 100 particles. The stellar {\tt SubDLe}  performs markedly better at identifying full substructures, with a higher $\text{M}_{\tt SF/SD}/\text{M}_{\tt SF}$. Average values for $\text{M}_{\tt SF/SD}/\text{M}_{\tt SF}$  are, indeed, 0.60 and 0.89, for the all-particles and stellar runs of   {\tt SubDLe}, respectively.
Having established the better performance of our stellar  {\tt SubDLe}, we keep on focusing only on star particles in this Section, though the general conclusions can be considered valid in both cases. We defer to future works for an equally detailed analysis of the all-particles version of {\tt SubDLe}, possibly trained in a framework with more computational resources available than the ones we had for this work.

In order to further test the performance of our stellar {\tt SubDLe}, we applied it to the identification of galaxies in 11 other simulated galaxy clusters at $z=0$ (identified as {\tt CL-[3-13]} in Table \ref{tab:clusters}), spanning the virial mass range $[10^{14},10^{15}] \ \text{M}_{\odot}$. Based on the previous analysis, {\tt SubDLe} has most difficulty in recovering large (in comparison with the d-grid size) substructures and in separating close (in comparison with the d-grid resolution) ones. 
While the former does not compromise the   {\tt SubDLe}'s performance as a peak locator, the latter does. We expect galaxies in clusters to be closer on average than their counterparts in the field or in small groups, as per definition of cluster of galaxies. Thus the cluster environment that we are probing in this work is the most challenging scenario for   {\tt SubDLe}. 

The cluster analysed so far, {\tt CL-2}, is one of the smallest of the sample; to analyze our most massive clusters, we had to deal with a significant memory occupation of the d-grids. For this reason, when necessary, we modified our sampling strategy of the cluster region, with overlapping d-grids to avoid missing information at the borders, opting instead for a sort of sparse sampling. To this purpose, we built coarse d-grids, each encompassing the whole cluster, and thus spanning regions of few Mpcs per side (the exact size depending on the cluster), with a resolution of 640 kpc. We then defined a threshold $\text{M}_{\rm th}$ for the mass assigned to each of those coarse pixels through the NGP scheme, below which the pixel is excluded from this analysis. In pixels assigned with a mass $\text{M}_{\rm NGP}>\text{M}_{\rm th}$, a $64^3$ sub-d-grid is built with 10 kpc resolution. In this analysis, the extension of the clusters and the memory resources allowed us to set $\text{M}_{\rm th}=0$, thus only empty regions are excluded. On the other hand, this simple scheme does not allow us to analyze a given (higher-resolution) d-grid of the cluster more than once, as in the previous case with overlapping d-grids. 

\begin{figure}
\centering
\includegraphics[width=0.475\textwidth]{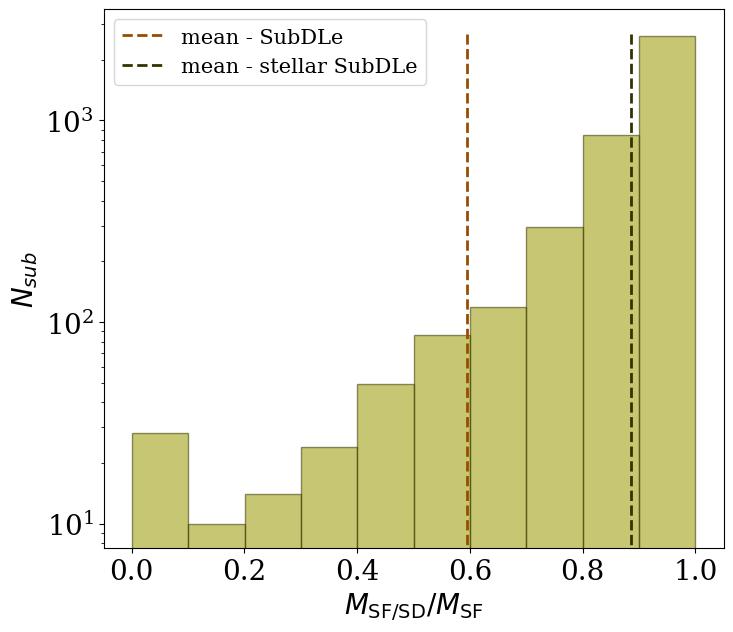}
\caption{\label{fig:mratio_allstars} Distribution of the ratio $\text{M}_{\tt SF/SD}/\text{M}_{\tt SF}$  in all galaxies identified by our stellar {\tt SubDLe}  and matched to a {\tt SubFind} substructure for all the 12 clusters analysed at $z=0$. The mean value is traced by the black dashed line, while the orange dashed line represents the mean value for the all-particle run of {\tt SubDLe} on cluster {\tt CL-2} (associated to the orange histogram shown in Fig. \ref{fig:histogramd4}).}
\end{figure}

Fig. \ref{fig:mratio_allstars} shows the results for all clusters analysed at $z=0$ ({\tt CL-[2-13]}) in terms of $\text{M}_{\tt SF/SD}/\text{M}_{\tt SF}$; the distribution is consistent with that shown for the single cluster {\tt CL-2} in Fig. \ref{fig:histogramd4}, with the same mean value of 0.89 (black dashed line, note that the chosen y-scale here is logarithmic, due to the much higher number of galaxies when considering 12 clusters), significantly larger than the mean value for the all-particles {\tt SubDLe} run (orange dashed line), 0.60.

Table \ref{tab:furthertests} reports the percentages of  galaxies identified by {\tt SubFind}, which are detected by our stellar {\tt SubDLe} (\emph{completeness}) in the 11 clusters {\tt CL-[3-13]}, along with the percentage of the stellar {\tt SubDLe} identified galaxies which have a {\tt SubFind} counterparts (\emph{purity}), following our matching criterion, and the wall-clock time requested by {\tt SubDLe} runs. The latter includes the actual propagation of d-grids through the network to recover the s-grids, but does not take into account the time required for building the d-grids, which might be already available during the evolution of a simulation (e.g. if gravity is solved through a Particle-Mesh approach or if a Eulerian hydrodynamical scheme is adopted) for stars, nor the final FoF run on pixels, which should be negligible anyway due to the relatively small number of active pixels in each s-grid.
The average completeness of our stellar   {\tt SubDLe}, weighted by the number of resolved galaxies in each cluster (reported in the fourth column of Table \ref{tab:clusters}), is $82$ per cent in the set of 12 clusters at $z=0$ (including {\tt CL-2}), while the weighted average purity is 0.88 (including again {\tt CL-2}, whose identification reported a purity of 0.82). We have already partially discussed the causes of the loss in completeness and will focus further on this aspect later in this Section. The purity is sufficiently high, even higher than the completeness, meaning that the bulk of {\tt SubDLe} galaxies have been identified by {\tt SubFind}. {\tt SubFind} does not necessarily represent the ideal identifier, though we had to assume its catalogues as ground truth in the framework of a ``supervised'' learning. Nevertheless we briefly investigated the nature of the ``spurious'' identification (in comparison with {\tt SubFind} resolved substructures) of {\tt SubDLe} galaxies. We relaxed the resolution limit for {\tt SubFind} galaxies to a minimum of 20 star particles and repeated the stellar   {\tt SubDLe}/{\tt SubFind} matching process, gaining an increase in weighted average purity from 0.88 to 0.96, confirming that almost all of the {\tt SubDLe} galaxies correspond to gravitationally bound structures found by {\tt SubFind}. At the same time, the completeness has decreased from 0.82 to 0.65, as expected since our stellar {\tt SubDLe} might very well be able to identify substructures of very few stars where there is close to zero background, while this will be harder to do in the denser central regions of the clusters. We will focus on how the position in the cluster affects the completeness later in this Section.  

\begin{table}
	\centering
	\caption{Results of identification of galaxies in 11 simulated galaxy clusters analyzed with our stellar {\tt SubDLe}; the first column reports the cluster ID, the second and third ones show the completeness and purity of our stellar {\tt SubDLe}  catalogues, in comparison with {\tt SubFind}, and the fourth one reports the wall clock time (in seconds) taken by {\tt SubDLe} for the processing of the d-grids.}
	\label{tab:furthertests}
	\begin{tabular}{lccr} 
		\hline
		ID & Completeness & Purity & {\tt SubDLe} time \\
           & & & [s] \\
		\hline
		{\tt CL-3} & 0.80 & 0.89 & 53\\
		{\tt CL-4} & 0.78 & 0.86 & 21\\
		{\tt CL-5} & 0.88 & 0.82 & 53\\
		{\tt CL-6} & 0.84 & 0.88 & 41\\
      	{\tt CL-7} & 0.75 & 0.65 & 74\\
        {\tt CL-8} & 0.81 & 0.86 & 105\\
        {\tt CL-9} & 0.78 & 0.91 & 49\\	
        {\tt CL-10} & 0.84 & 0.92 & 146\\		
        {\tt CL-11} & 0.82 & 0.86 & 36\\
        {\tt CL-12} & 0.83 & 0.90 & 152\\
        {\tt CL-13} & 0.79 & 0.82 & 59\\
		\hline
	\end{tabular}
\end{table}

The wall-clock time taken by {\tt SubDLe} to perform galaxy identification in these clusters, shown in the fourth column of Table \ref{tab:furthertests} is quite interesting to assess whether {\tt SubDLe} is suitable for frequent on-the-fly identifications of galaxies in cosmological simulations. With the above mentioned caveats about possible overheads, all the analyses required less than 3 minutes per cluster, which represents a remarkable improvement with respect to {\tt SubFind}. For reference, running {\tt SubFind} on one of such clusters required about half an hour when occupying one node of the Marconi-100 machine at the CINECA Supercomputing Center\footnote{https://wiki.u-gov.it/confluence/pages/viewpage.action?pageId=336727645}.

\begin{figure}
 \centering
 \begin{subfigure}[a]{0.45\textwidth}
   \includegraphics[width=\textwidth]{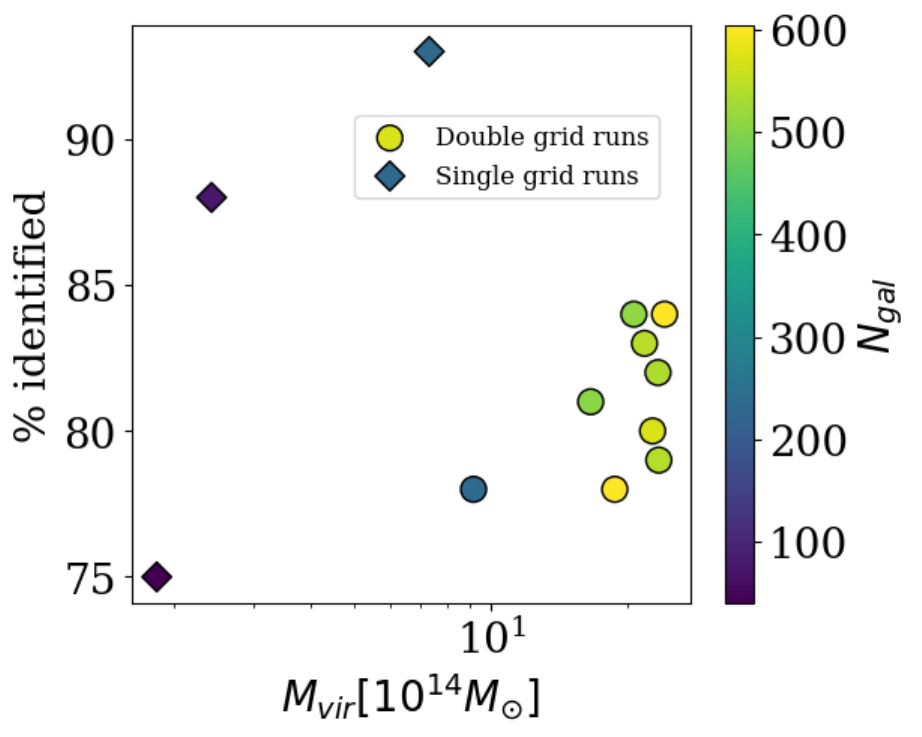}
   \label{fig:Mvir}
 \end{subfigure}
 \begin{subfigure}[b]{0.45\textwidth}
  \includegraphics[width=\textwidth]{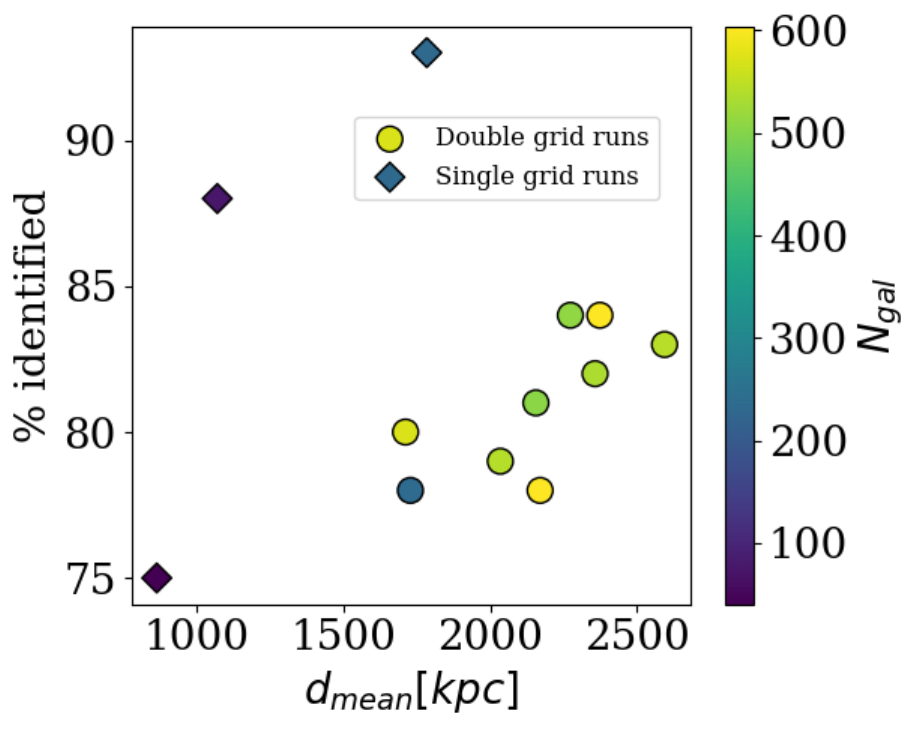}
  \label{fig:Dmedian}
 \end{subfigure}
 \begin{subfigure}[c]{0.45\textwidth}
  \includegraphics[width=\textwidth]{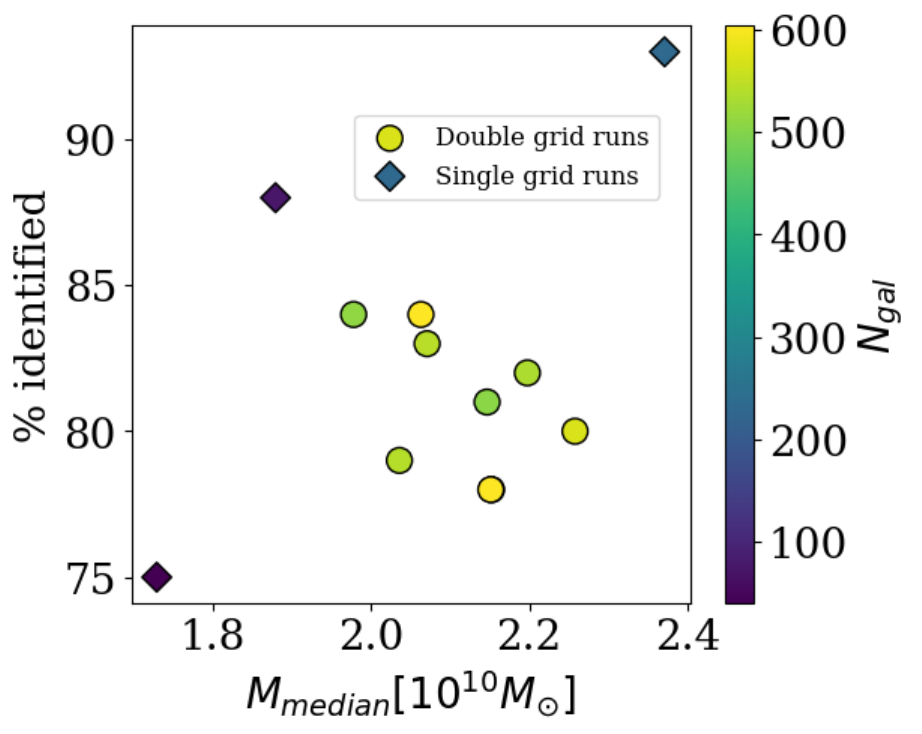}
  \label{fig:Mmedian}
 \end{subfigure}
 \caption{Scaling of the percentage of resolved {\tt SubFind} galaxies in the 12 clusters analysed at $z=0$ ({\tt CL-[2-13]}), identified by our stellar {\tt SubDLe}, with cluster properties: total virial mass (upper panel), median inter-galaxy separation within one virial radius (middle panel) and median galaxy stellar mass (lower panel), color-coded with the total number of {\tt SubFind} galaxies (the ones we consider resolved enough, with at least 100 star particles).}
 \label{fig:clustproperties}
\end{figure}

\subsection{Testing performance against galaxy properties and environment}
For a better understanding of the capability of our stellar {\tt SubDLe} , we investigate what drives the cluster-by-cluster variations of the completeness of {\tt SubDLe} galaxy catalogues, looking for trends with cluster properties. In this analysis, we include also cluster {\tt CL-2}, which yielded the highest (93 per cent) percentage of identified resolved galaxies. This higher value might be due to the fact that {\tt CL-2} has very similar properties to {\tt CL-1}, which was used for the training of {\tt SubDLe}, pointing towards a slight over-fitting of the model. To verify this, we look into possible correlations between the completeness of the stellar {\tt SubDLe} catalogues and the properties of clusters in Fig. \ref{fig:clustproperties}. This shows the completeness of the identification in the 12 clusters at $z=0$ as a function of the total virial mass (upper panel), the median inter-galaxy distance for {\tt SubFind} galaxies within the virial radius (middle panel) and the median galaxy stellar mass (lower panel). The circles in the plot represent clusters that required a treatment with a second hierarchy of sparsely distributed sub-grids (with $\text{M}_{\rm th}=0$, as described above) and the diamonds represent the three smaller clusters that did not require such treatment to fit into the memory constraints. The clusters are color-coded according to the total number of {\tt SubFind} galaxies. We note that clusters with a lower statistics of galaxies (the three diamonds, {\tt CL-[2,5,7]}, and the darker green circle, {\tt CL-4}) appear to be more randomly scattered (thus disproving the possibility of over-fitting), with an average increasing completeness with larger cluster mass, while the clusters with a larger number of galaxies follow a tighter locus with a stronger dependence upon the total virial mass of the clusters, suggesting that the identification is less complete in less massive clusters. This might be due to a larger fraction of well resolved galaxies in the most massive systems. These clusters, though, span a small range of virial masses (see Table \ref{tab:clusters}), so it is difficult to draw firm conclusions on the environment dependence of the model performance from this set of simulated clusters.
There is also an apparent correlation with the mean inter-galaxy distance, with two of the diamonds ({\tt CL-2} and {\tt CL-5}) acting as outliers; nevertheless, we do not trust the statistical significance of these two clusters to define a distinct locus of higher completeness for clusters identified with a different spatial sampling, in consideration also of the low completeness of {\tt CL-7} (0.75), which was analysed in the same way. The general trend of increasing completeness with increasing inter-galaxy separation can be understood as the larger the separation, the lower the probability for {\tt SubDLe} to merge two substructures and thus reducing the completeness. On the other hand, the weak apparent anti-correlation with the median galaxy stellar mass shown in the bottom panel (here {\tt CL-4} is not visible because hidden behind one the yellow circles) is less intuitive. We interpret this as an indirect effect of the correlation shown in the middle panel, since a larger fraction of slightly more massive galaxies might produce a larger fraction of close galaxies on their way to an actual merger, thus with a small separation.  

\begin{figure*}
    
    \begin{subfigure}{0.475\textwidth}
        \includegraphics[width=\textwidth]{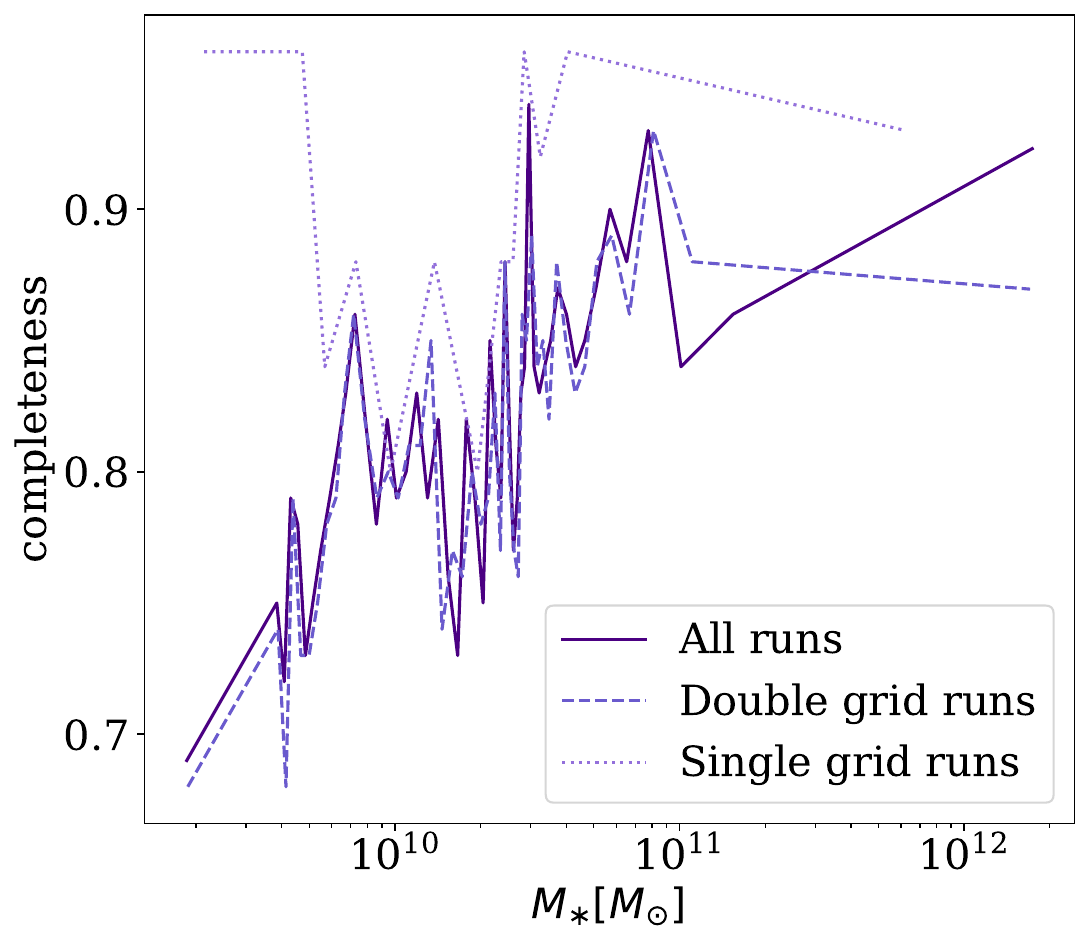}
    \end{subfigure}
    \hfill
    \begin{subfigure}{0.475\textwidth}
        \includegraphics[width=\textwidth]{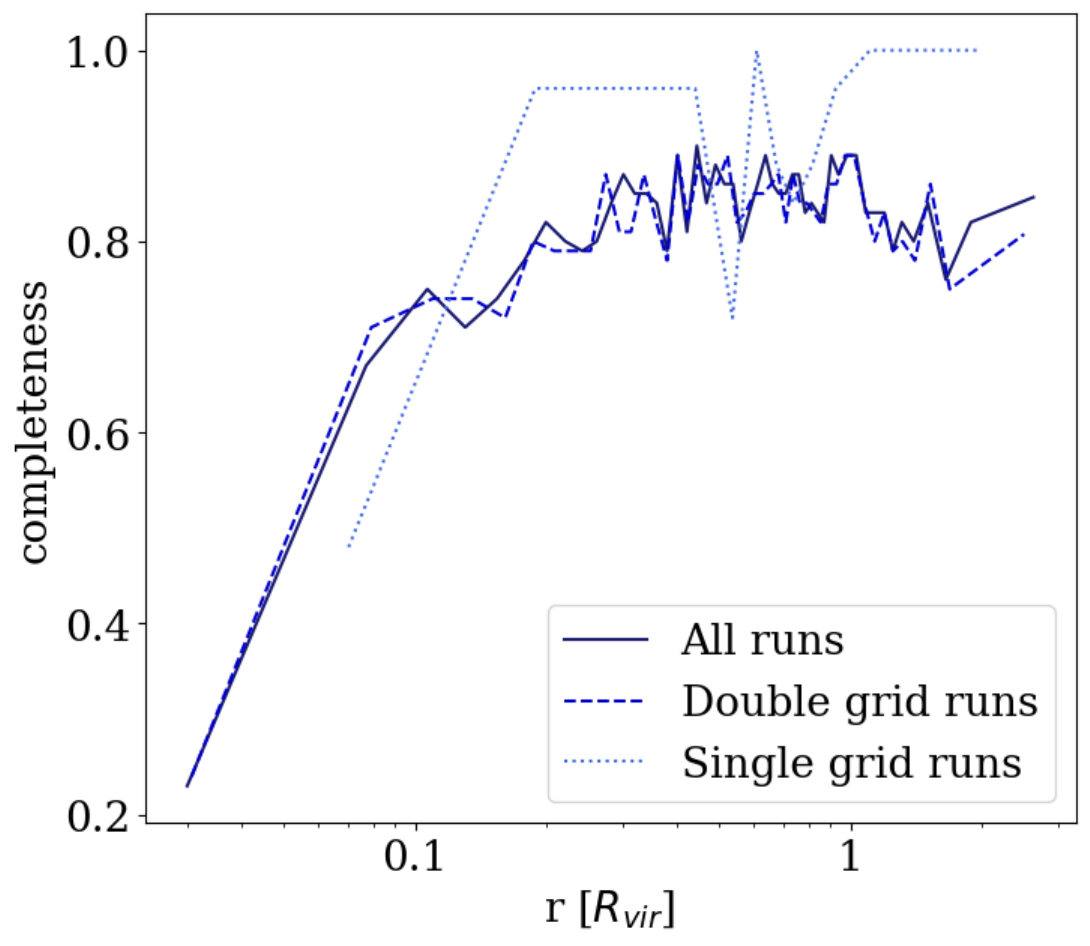}
    \end{subfigure}
    \vskip\baselineskip
    \begin{subfigure}{0.475\textwidth}
        \includegraphics[width=\textwidth]{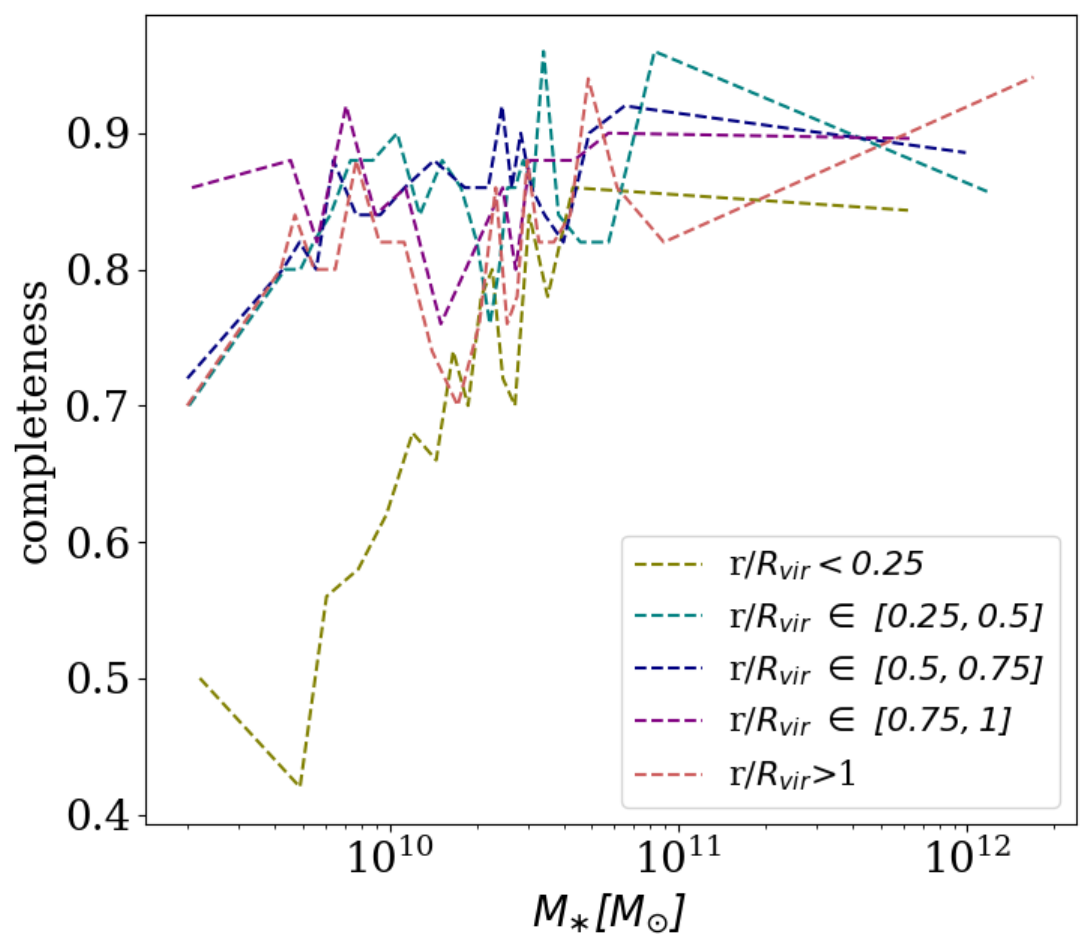}
    \end{subfigure}
    \hfill
    \begin{subfigure}{0.475\textwidth}
        \includegraphics[width=\textwidth]{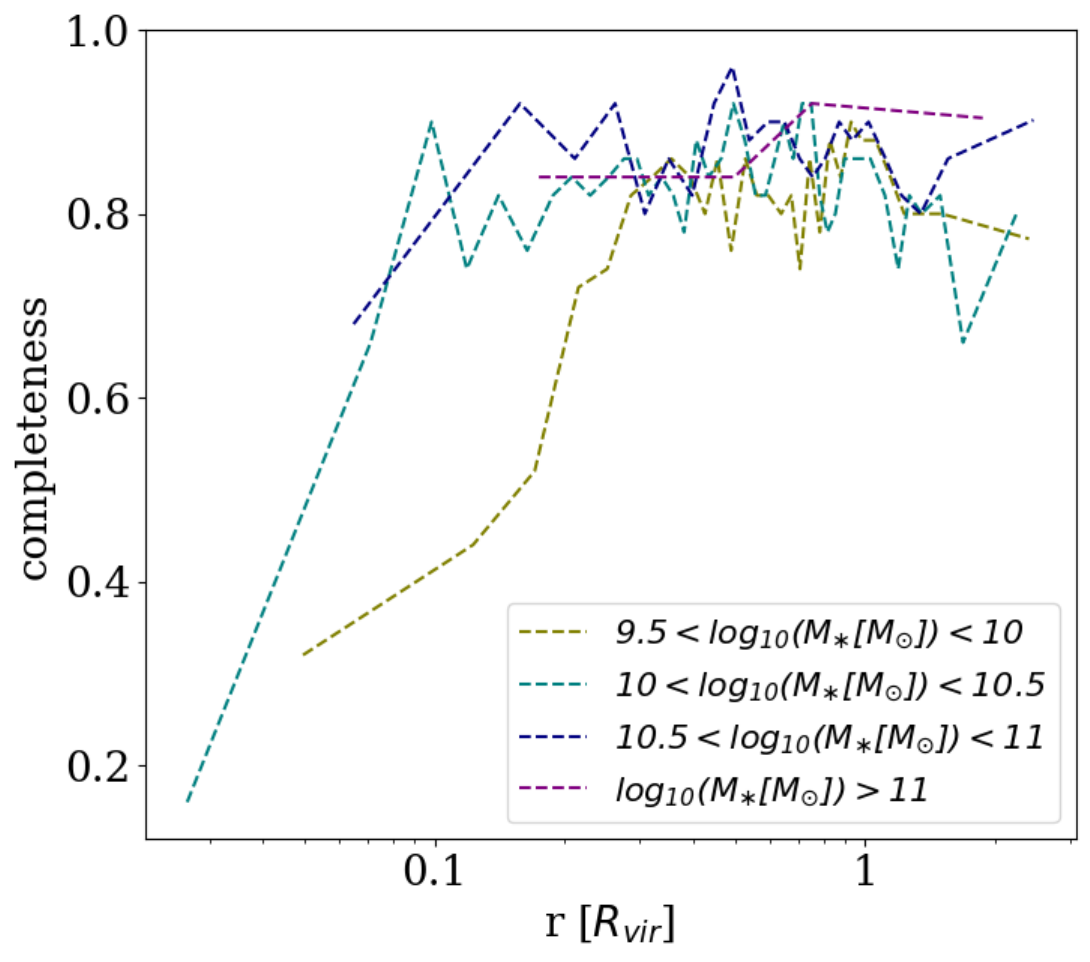}
    \end{subfigure}
    \caption{
 Dependence of the stellar {\tt SubDLe} \emph{completeness} (defined as the fraction of {\tt SubFind} galaxies identified by our stellar {\tt SubDLe}) on {\tt SubFind} stellar masses (left panels) and 3D distance from the cluster center (in units of virial radii, right panels), for all the clusters analysed at $z=0$ (see Table \ref{tab:furthertests}). In the upper panels we show, with solid lines, the results for all the analysed clusters, while dashed and dotted lines represent clusters analysed with a single grid and a double grid, respectively (see text). In the lower left panel different colours of the dashed curves correspond to different radial ranges, as indicated in the legend. In the lower right panel different colors correspond instead to different intervals of the stellar mass of the galaxies, as indicated in the legend. 
    }
    \label{fig:completeness_rm}
\end{figure*}

One further step can be done to isolate the factors affecting the model performance, by studying the completeness as a function of {\tt SubFind} stellar masses and distance from the center of the cluster (defined as the position of the most bound particle in the halo). This is shown in Fig. \ref{fig:completeness_rm}; the top left panel shows the percentage of {\tt SubFind} galaxies identified by our stellar {\tt SubDLe} as a function  of {\tt SubFind} stellar mass, for all the clusters analysed at $z=0$ ({\tt CL-[2-13]}), in solid line, for the ones analysed with a double grid (with $\text{M}_{\rm th}=0$, {\tt CL-3-4,6,8-13}), in dashed line, and the ones analysed with a single grid, in dotted line. As already discussed, the latter have low statistics, with a total of 346 galaxies (see Table \ref{tab:clusters}). The completeness, for all clusters, increases with stellar mass, with an apparent difference below $\approx 6 \cdot 10^9 \ \text{M}_{\odot}$ for galaxies belonging to clusters processed with a single grid. Owing to the relatively poor statistics in this sample and the proximity to the resolution limit, we do not deem this discrepancy significant. The increasing completeness with stellar mass is in line with the expectation that the most massive galaxies are better resolved against the background. The top right panel shows the completeness as a function of the cluster-centric radius, expressed in units of the virial radius, with the same distinction between different lines adopted in the top left panel. As expected, we see an improvement in performance the farther the galaxies are from the over-crowded central regions of the clusters, up to $\approx 0.2-0.3 \ \text{R}_{\text{vir}}$, followed by a saturation at $\sim 80$ per cent completeness in the external regions. The dependence of completeness on the distance from the center is much stronger than what is observed with the stellar mass (see the different scales in the y-axis of the top left and top right panels), at least in the innermost regions, and is ubiquitous among galaxies processed in different ways. 

In any case, a larger sample of simulations uniformly sampling the halo mass range from groups to rich clusters would be required to draw a picture of how the external environment of galaxies affects the stellar   {\tt SubDLe}'s performance.
As a further step, we tried to disentangle the dependence on the stellar mass and on the distance from the center of the cluster, by separating galaxies in radial and stellar mass bins (bottom panels of Fig. \ref{fig:completeness_rm}). The bottom left panel shows the completeness of identification of galaxies in 5 radial bins, as a function of stellar mass. The bottom right panel shows the completeness in 4 stellar mass bins, as a function of distance from the cluster center. Here, in order to provide a sufficient number of galaxies in each bin, we included all the clusters analysed at $z=0$. The separation in radial bins shows that the stellar-mass dependence is the strongest in the most central region (below 0.25 $\text{R}_{ \text{vir}}$), where the overdense environment affect more significantly the identification of the smaller galaxies. This is confirmed by the bottom right panel, where we  see an approximate hierarchy in the curves describing the completeness as a function of cluster-centric radius for different mass bins. In the central region of the clusters (below $\approx 0.25 \ \text{R}_{ \text{vir}}$), the completeness increases with the distance from the cluster center and the curves representing less massive galaxies are lower (apart from statistical fluctuations). In conclusion, the main limitation of our stellar {\tt SubDLe} is in the identification of small galaxies within the crowded core regions of galaxy clusters. This is expected in our algorithm and can be attenuated by increasing the resolution of the grids.

\begin{figure}
    \centering
    \includegraphics[width=0.475\textwidth]{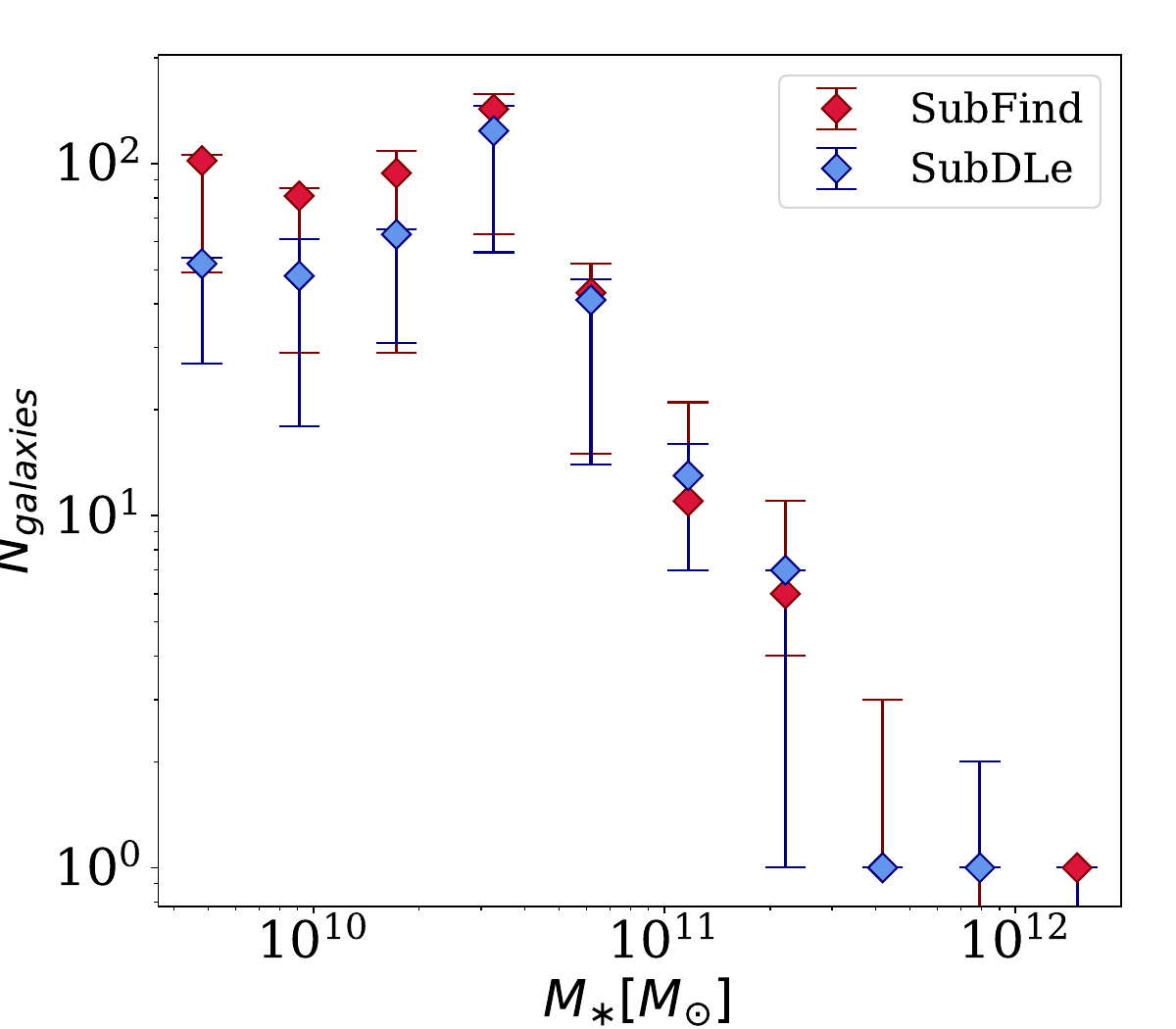}
    \caption{ \label{fig:gsmf} Comparison of the distributions of stellar masses for our simulated set of galaxy clusters at $z=0$, as predicted by the subhalo catalogues produced by {\tt SubFind} (red diamonds) and by our stellar {\tt SubDLe} (blue diamonds). The diamonds define the median counts in each stellar mass bin among the 12 different clusters, while the errorbars mark the 25th/75th percentiles.}
\end{figure}

Finally, we compare the distribution of stellar masses produced by {\tt SubFind} and by the stellar {\tt SubDLe} in clusters at $z=0$ in Fig. \ref{fig:gsmf}. The diamonds mark the median counts among different clusters and the errorbars define the 25th/75th percentiles, blue for {\tt SubDLe} and red for {\tt SubFind}. In the low-mass end where the resolution limit of the simulation dominates the shape of the distribution, the {\tt SubDLe} stellar mass distribution is slightly lower due to missing identifications and {\tt SubDLe} galaxies being smaller than {\tt SubFind}'s. In the  well resolved high-mass end, though, there is agreement in shape and normalization among the two distributions.

\subsection{Testing the robustness of the stellar   {\tt SubDLe} }
We now focus on testing the robustness of the stellar {\tt SubDLe} on the galaxy cluster {\tt CL-13} at redshift $z=1$, with the training having been carried out at $z=0$. At $z=1$ this cluster has a virial mass $\text{M}_{\text{vir}} \approx 7 \times 10^{14} \text{M}_{\odot}$. In principle a higher degree of merging and more dynamically disturbed clusters are expected at $z=1$, which could impact the performance of {\tt SubDLe} in identifying galaxies. This complements the previous tests in assessing the capability of the model to extrapolate outside the training framework, which is always a primary concern in developing Machine Learning models. Here, with the uniform sampling of the cluster volume, we extracted 1287 $64^3$ d-grids with a resolution of 10 physical kpc. The percentage of resolved galaxies identified is 77 per cent, consistent with the 79 per cent of completeness obtained at $z=0$ for the same cluster and with the average performance achieved by stellar {\tt SubDLe} with a 10 kpc resolution at $z=0$. This confirms the robustness of the method also when different evolutionary stages of galaxy clusters are considered.

As a last test, we assess the performance of the stellar {\tt SubDLe} on {\tt CL-13} at $z=1$, by applying a higher resolution d-grids. Our stellar {\tt SubDLe} was trained on galaxies of different sizes, so we expect it to be able to identify galaxies even when using a higher-resolution d-grid, with respect to which their sizes will be larger. In general, we expect our method to be able to identify galaxies that were already identified at lower resolution. To increase resolution, we chose a grid size of 2 kpc and, due to memory occupation, perform a sparse sampling of the cluster volume as described before, still keeping $\text{M}_{\rm thr}=0$.
By doing this, we expect to: \emph{(i)} increase the sensitivity of the network to small substructures; \emph{(ii)} separate composite systems made up of close galaxies artificially merged by {\tt SubDLe} due to the poor relative resolution of d-grids in the denser regions. In general, we expect that the higher resolution d-grid will increase the percentage of {\tt SubFind} resolved galaxies found by our stellar {\tt SubDLe}. On the other hand, by increasing the resolution, the average mass per grid pixel decreases, resulting in a more noisy discontinuous NGP density assignment on the grid. This in turn might impact the model's performance since CNNs are efficient on input data with strong spatial dependencies. Moreover, by reducing the pixel size, while keeping fixed the number of pixels in each grid, the whole d-grid encompasses a smaller volume ($128 \ {\rm kpc}$ per side), which we expect to have a negative impact on the recovered spatial extension of the substructures.

\begin{figure*}
\centering
    \begin{subfigure}{0.475\textwidth}
        \includegraphics[width=\textwidth]{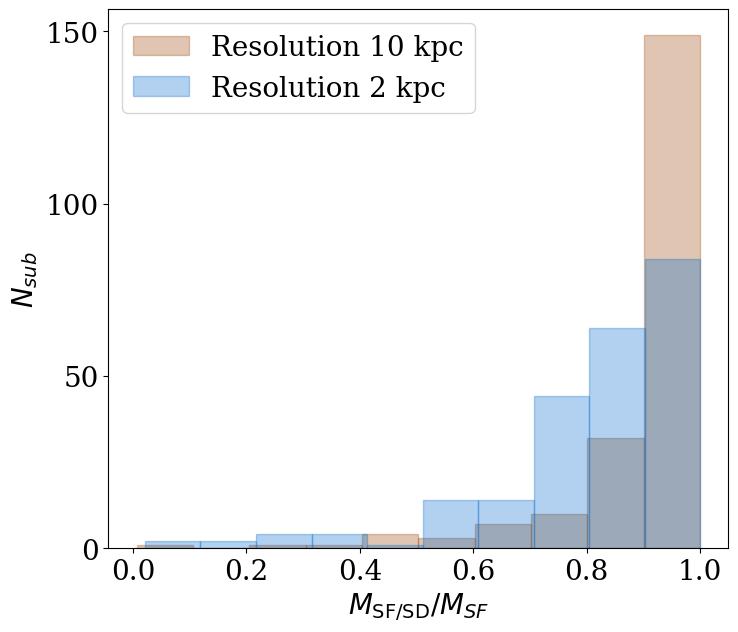}
    \end{subfigure}
     \begin{subfigure}{0.475\textwidth}
        \includegraphics[width=\textwidth]{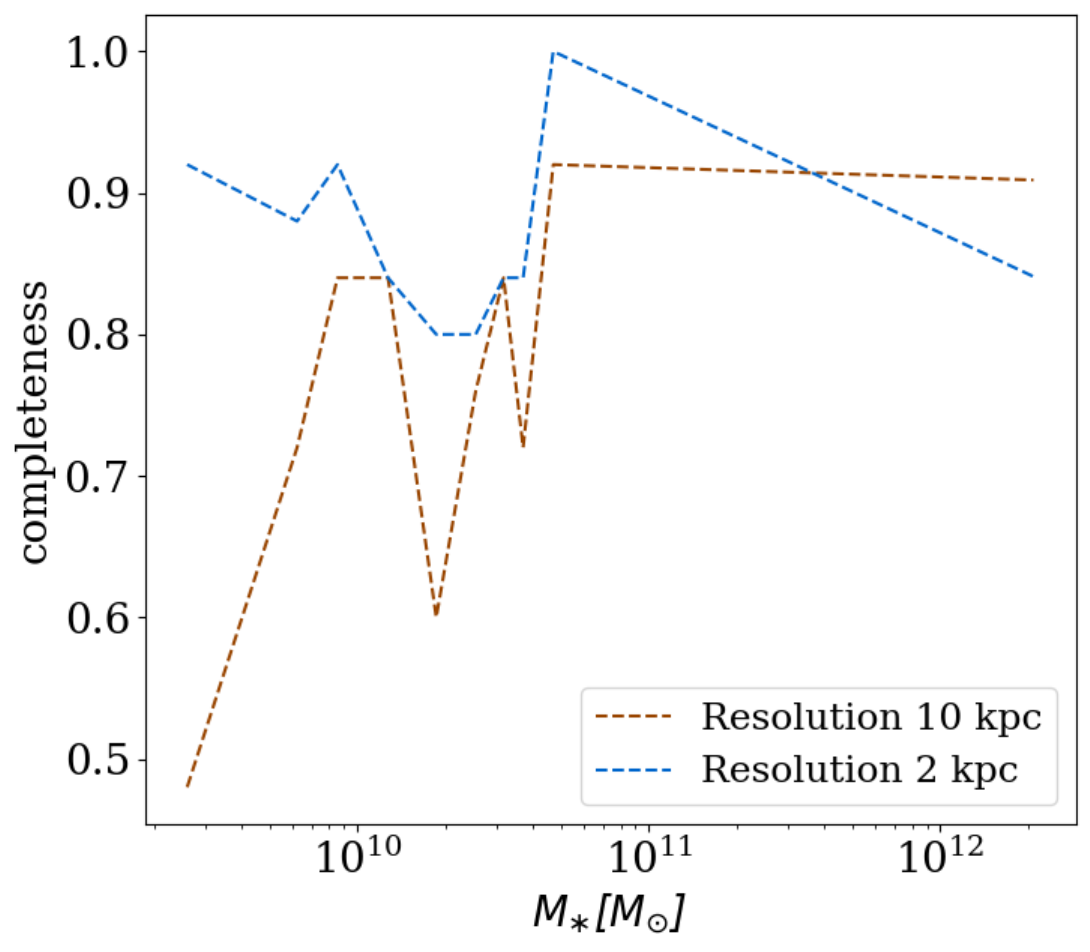}
    \end{subfigure}
    \caption{\label{fig:D24threshold} \emph{Left panel}: Distributions of values of $\text{M}_{\tt SF/SD}/\text{M}_{\tt SF}$  for {\tt SubDLe} galaxies matched to {\tt SubFind} galaxies. Results are shown for the cluster {\tt CL-13} at $z=1$, for {\tt SubDLe} runs with the standard resolution of 10 kpc (orange) and with the higher resolution of 2 kpc (blue). \emph{Right panel}: stellar {\tt SubDLe} completeness as a function of stellar mass in cluster {\tt CL-13} at $z=1$, for 2 kpc (blue) and 10 kpc (orange) resolution.}
\end{figure*}

The results of this test are shown in Fig.\ref{fig:D24threshold}. In the left panel we show the distribution of $\text{M}_{\tt SF/SD}/\text{M}_{\tt SF}$  for {\tt SubFind}/stellar {\tt SubDLe} matches, for the run at lower (10 kpc, orange histogram) and higher (2 kpc, blue histogram) resolution. The average values vary from 0.9 at 10 kpc resolution (consistent with the mean value at $z=0$), to 0.7 at 2 kpc resolution. As expected, the completeness of the galaxy mass distribution recovered by the 2 kpc resolution run of the stellar {\tt SubDLe} is slightly lower than for the 10 kpc resolution.
The most interesting aspect is though the increased percentage of identified peaks in the stellar mass density distribution when running our stellar {\tt SubDLe} at higher resolution, which proves that the model can reach peak performance by simply increasing the resolution. In fact, the percentage of {\tt SubFind} resolved galaxies identified by the stellar {\tt SubDLe} on cluster {\tt CL-13} at $z=1$ increased from 77 to 87 per cent when increasing the grid resolution. This is shown in the right panel where we plot the completeness as a function of stellar mass (blue and orange for 2 and 10 kpc resolution, respectively). This Figure highlights that the higher resolution adopted in the d-grids boosts the performance of {\tt SubDLe} mainly by recovering a larger fraction of small galaxies, the ones that {\tt SubDLe} fails to disentangle from the background and other close substructures in the central region of clusters at lower resolution. This further demonstrates the capability of our model in the identification of substructures in the simulated distribution of stars. 

\subsection{Testing with a larger training set}
\label{sec:2ndtraining}
As already emphasized, we expect an improved performance from {\tt SubDLe} once it is trained on a much larger data set, and using more performant computing nodes. In order to explicitely show that we are not saturating the model's performance, we carried out a training on a data set which is twice as large as the original one, taken from clusters {\tt CL-3} and {\tt CL-8}, which is still feasible in our computing framework. Then, we tested on the remaining clusters at $z=0$ and compared the results with the previously trained model, for those clusters which are not part of the training data set in neither models.

\begin{figure}
    \centering
    \includegraphics[width=0.475\textwidth]{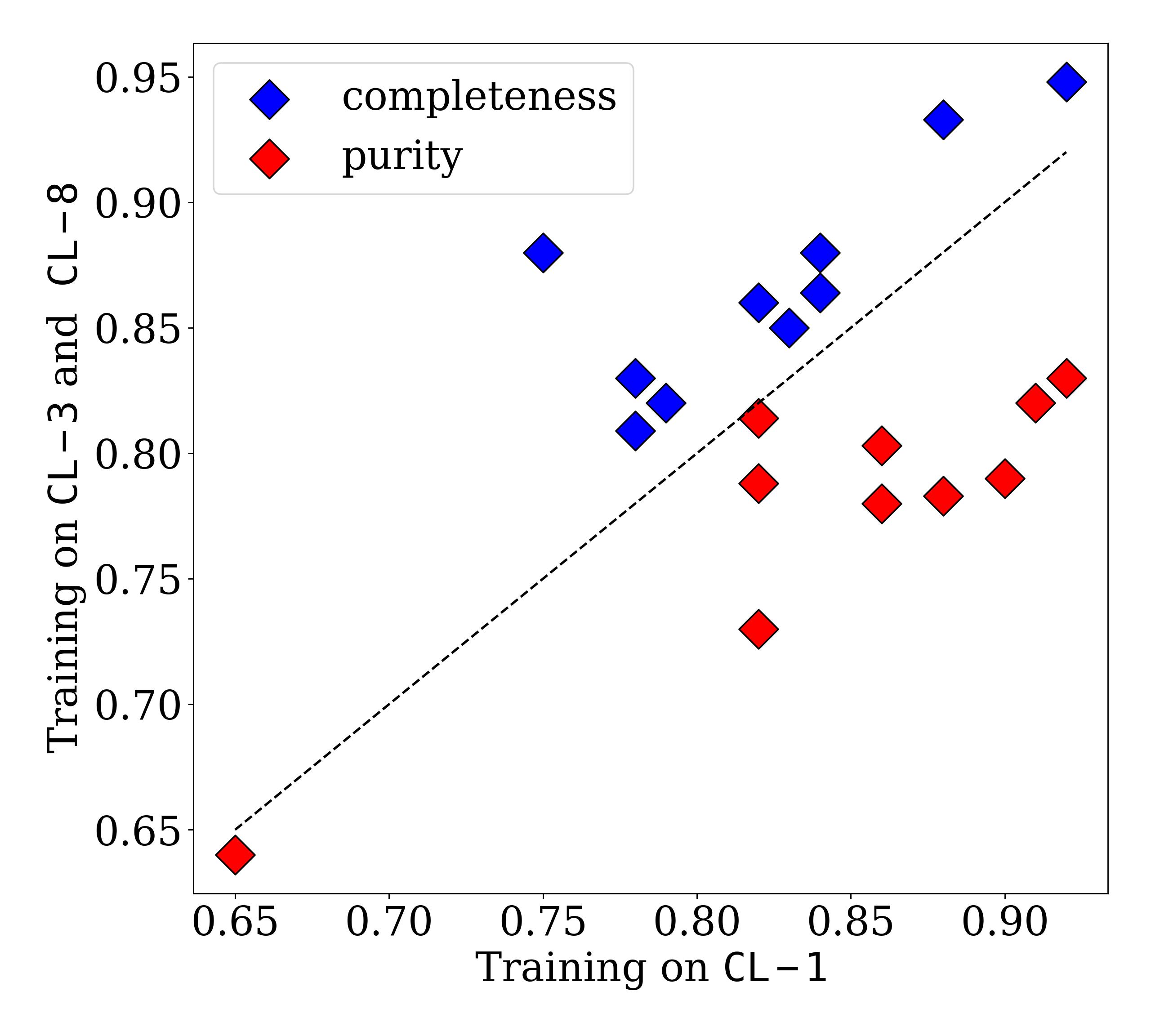}
    \caption{\label{fig:2ndtraining} Completeness (blue diamonds) and purity (red diamonds) of {\tt SubDLe} identifications, referred to {\tt SubFind} catalogues, for galaxies in $z=0$ clusters. The y-axis refers to metrics calculated on the catalogs given by the model trained on two clusters ({\tt CL-3} and {\tt CL-8}), while the x-axis refers to results of the original stellar {\tt SubDLe} discussed in the previous sections.}
\end{figure}

The results of this test are summarized in Figure \ref{fig:2ndtraining}, where completeness and purity reached by the newly trained model are shown for each test cluster, against the corresponding quantities obtained from the version of the model trained on a single cluster. We note that there is a systematic improvement in completeness (blue diamonds), with the overall completeness (weighted with the number of resolved galaxies in each cluster) increasing from 0.82 to 0.86. On the other hand, the purity is systematically lower, meaning that the new version of {\tt SubDLe} identifies more galaxies which are missed by {\tt SubFind}. This improvement is quite promising in the perspective of a future use of a more extended training data set, spanning possibly different redshifts, implementation of the galaxy formation model and numerical resolutions, which we aim at carrying out in future developments of this work.

\section{Conclusions}
\label{sec:conclusions}
In this work, we have presented the first application of deep learning techniques to perform identification of substructures in cosmological hydrodynamical simulations, with the goal of providing a much faster alternative to procedural algorithms typically used for this purpose (see Section \ref{sec:introduction} for a brief review of some popular subhalo identifiers). Substructure identification represents a fundamental step to extract useful information from the raw output of cosmological hydrodynamical simulations and is typically performed by algorithms that, with different  approaches, search for groups of gravitationally bound particles at the peaks of the density field sampled by the simulations, within a larger structure. This is computationally costly and results in long times-to-solution that are unfit for frequent identifications of galaxies during the evolution of simulations. The on-the-fly identifications of substructures with a fine time cadence, on the other hand, could have interesting applications in the implementation of sub-resolution models of, e.g., star formation and feedback, when they are parametrized on global properties of galaxies, such as their stellar masses.  For example, having access to the information on stellar masses of galaxies frequently during the evolution of the simulation would allow to associate a BH mass to each galaxy according to the Magorrian relation \citep{magorrian}.

Deep learning, and Machine Learning in general, consists in statistical methods to approximate solutions to many classes of problems and offers the advantage of much shorter computational times. We borrowed the architecture of an efficient Fully Convolutional Network, U-Net \citep[][]{unet}, developed for pixel-by-pixel identification of objects in 2D images, and extended it to the task of identifying 3D substructures in simulations. In this work we presented {\tt SubDLe}, a substructure finder which incorporates this 3D generalization of U-Net. We analysed the {\tt DIANOGA} simulations (see Section \ref{sec:methods}), a set of zoom-in cosmological hydrodynamical re-simulations \citep[][]{tormen97} of galaxy clusters, which include a sub-resolution model of star formation from a multi-phase interstellar medium, as well as stellar and AGN feedback. As the ground truth on which to train {\tt SubDLe}, we used the subhalo catalogues extracted from these simulations by the subhalo finder {\tt SubFind} \citep[][]{subfind,stellarsubfind}. We applied our {\tt SubDLe} to 3D Nearest Grid Point mass density grids extracted from the set of simulations, splitting the whole cluster regions in 640 kpc-per-side sub-grids with 10 kpc resolution (\emph{d-grids}). Twin grids (\emph{s-grids}) reporting a label `1' at grid meshes containing a majority of substructure-bound particles (and `0' elsewhere) were also produced for the training, based on {\tt SubFind}'s catalogues.

We trained our {\tt SubDLe} on a NVidia Tesla P100 GPU on the Kaggle platform on a large number of representative d-grids. The test on a different cluster, after matching {\tt SubDLe} and {\tt SubFind} subhalos, resulted in a completeness of the output subhalo catalogues of 0.80 with respect to {\tt SubFind}. We then focused on the identification of galaxies, defined as concentration of star particles. As such, galaxies provide much more contrasted substructures against the background field, thus facilitating their identification performed by {\tt SubDLe}. We called this newly trained version of the neural network  ``stellar   {\tt SubDLe}'', to distinguish it from the previous version based on the density traced by all particles. 
The same test cluster analysed with the standard {\tt SubDLe} yielded a completeness of 0.93 with stellar {\tt SubDLe}. Furthermore, averaging over a sample of 12 simulated clusters at $z=0$ we reached 0.82 in completeness and an even higher purity of 0.88 (though most of the ``spurious'' identifications actually correspond to under-resolved {\tt SubFind} galaxies which were dropped from the input catalogue to make it cleaner). The average {\tt SubDLe} time for processing one cluster is 72 seconds, much faster than the typical execution time of {\tt SubFind}, thus matching the requirement for short computational times. We note that in the estimate of the execution time we did not consider the time necessary to build mass density grids and the FoF run on grid meshes (see Section \ref{sec:methods}). We consider the FoF runs to add a negligible overhead, while mass grids are typically already built during a simulation for either gravity or hydrodynamics. 
\newline
We tested the robustness of our stellar {\tt SubDLe} by repeating the identification on one of the clusters analysed at $z=0$, at a different evolutionary stage, at $z=1$, obtaining a completeness comparable to the result at $z=0$. Furthermore, we tested the effect of increasing the resolution of the d-grids by reducing the grid pixel size from 10 kpc to 2 kpc. This led to an increase of the percentage of {\tt SubFind} galaxies that are identified by our method from 77 to 87 percent. This is due to the reduction of close substructures that were artificially merged together while adopting 10 kpc resolution grids. On the other side, the d-grids are 5 times smaller, which slightly impact the size of {\tt SubDLe} galaxies in comparison with the lower resolution runs.

We found that our stellar {\tt SubDLe} performs better than the all-particles version, not only in overall completeness, but also in reproducing the total mass of substructures. 
The most challenging aspect for our stellar {\tt SubDLe} is the identification of galaxies in the innermost part of the clusters. We found that the fraction of {\tt SubDLe} identified galaxies is decreasing towards the center of each cluster, within about 0.25 virial radii, especially for lower-mass galaxies (below few $10^{10} \ \text{M}_{\odot}$). The smaller galaxies are indeed easier to miss in the high-density cores of clusters, due to smaller separations and higher level of background density. As a consequence, we expect our stellar {\tt SubDLe} to perform even better in different environments (small groups and average fields).

In future developments of this study, we plan to perform a full assessment of how our stellar   {\tt SubDLe} performs in different environments, with different resolution and even different sub-grid physics, which might affect the distribution of star particles. As far as the improvement of the algorithm is concerned, we aim to further refine it by taking advantage of high-performance computing nodes, where multiple GPUs can access a larger shared memory. This would largely relax the constraints we had to respect in this exploratory study, in terms of size and resolution of the grid where the density field is reconstructed. This will allow us to test how increasing the physical extension of d-grids, while keeping the resolution fixed, can influence the mass distribution of the largest substructures which are cropped by our stellar   {\tt SubDLe}. We will be able to train on much larger data sets, which is expected to improve the performance of the model, as shown in the tests described in Section \ref{sec:2ndtraining}. We also plan on testing different mass assignment schemes, in order to enhance the contrast between substructures and background field in the training data sets, which we also expect to improve the mass distribution found by {\tt SubDLe} for each galaxy. As a further step, given the demonstrated improvement of completeness with increasing resolution of the d-grids, we plan to develop a multi-resolution adaptive implementation of {\tt SubDLe} , to zoom on higher-density regions in the simulation box and increase the resolution of the d-grids, while keeping the average regions at a reasonable resolution. This would require to access adaptively refined meshes to sample the mass density field in the simulation, which in many cases are already constructed by the gravity or by the hydrodynamic solvers of the simulation code. We will also test the use of multiple channels in the input grids, including dynamical information to help the separation of the artificially merged close substructures.

In conclusion, we have developed {\tt SubDLe}, a fast deep learning model capable of identifying a large fraction of galaxies in simulations of galaxy clusters, while we expect its performance to be even better within environments that are less crowded than galaxy clusters. The final step, besides its possible refinements mentioned above, would be to integrate it into a cosmological code and test it for highly frequent, on-the-fly galaxy identifications.  

\begin{acknowledgements}
      We thank Massimo Brescia and Ilaria Marini for useful discussions about Machine Learning, and an anonymous referee for constructive comments that helped improving the presentation of the results. We also thank the Kaggle platform for the access to their computing resources and  the platform at \url{http://www.figma.com} for several graphic resources reproduced in this work. 
      This paper is supported by: the Italian Research Center on High Performance Computing, Big Data and Quantum Computing (ICSC), project funded by European Union - NextGenerationEU - and National Recovery and Resilience Plan (NRRP) - Mission 4 Component 2, within the activities of Spoke 3, Astrophysics and Cosmos Observations; by the PRIN 2022 PNRR project (202259YAF) ``Space-based cosmology with Euclid: the role of High-Performance Computing''. We acknowledge partial financial support from the INFN Indark Grant and the INAF project ``CONNECTIONS'' (COllaboratioN oN codE development for future Cosmological simulaTIONS).
\end{acknowledgements}

\bibpunct{(}{)}{;}{a}{}{,} 
  
\bibliographystyle{aa}
\bibliography{refs}

\begin{appendix} 

\section{Basic structure of a CNN}
\label{sec:appendix}
In the typical application of CNNs for the analysis of images, each layer is characterized by a 3D input grid with a \emph{height}, a \emph{width} and a \emph{depth}; in the input grid, \emph{height} and \emph{width} represent the spatial dimensions, whereas \emph{depth} is the number of color channels. 

In the following, we refer to the input data of CNNs as images, using the common definition of pixel as an element of the 2D grid ($\emph{height} \times \emph{width}$), described by \emph{depth} values (the brightness in three color channels in the typical case of RGB images). 
The generalization from two to three spatial dimensions is straightforward, so we describe the 2D case to keep the notation simple.

\subsection{Convolution}
\label{sec:convolution}
A convolution involves an operation between a grid and a \emph{filter}, a small grid whose values are parameters to be optimized during the training stage.
Let us consider the $q$-th layer of a CNN with dimensions $L_q \times B_q \times d_q$ (meaning that the input grid has such dimensions in this layer) and a $F_q \times F_q \times d_q$  filter, where the depth $d_q$ is always the same as in the input grid. Usually $F_q \ll L_q$ and $F_q \ll B_q$, with typical values for $F_q$ being 3 or 5.\\ 

\begin{figure}
    \flushleft
    \includegraphics[width=0.5\textwidth]{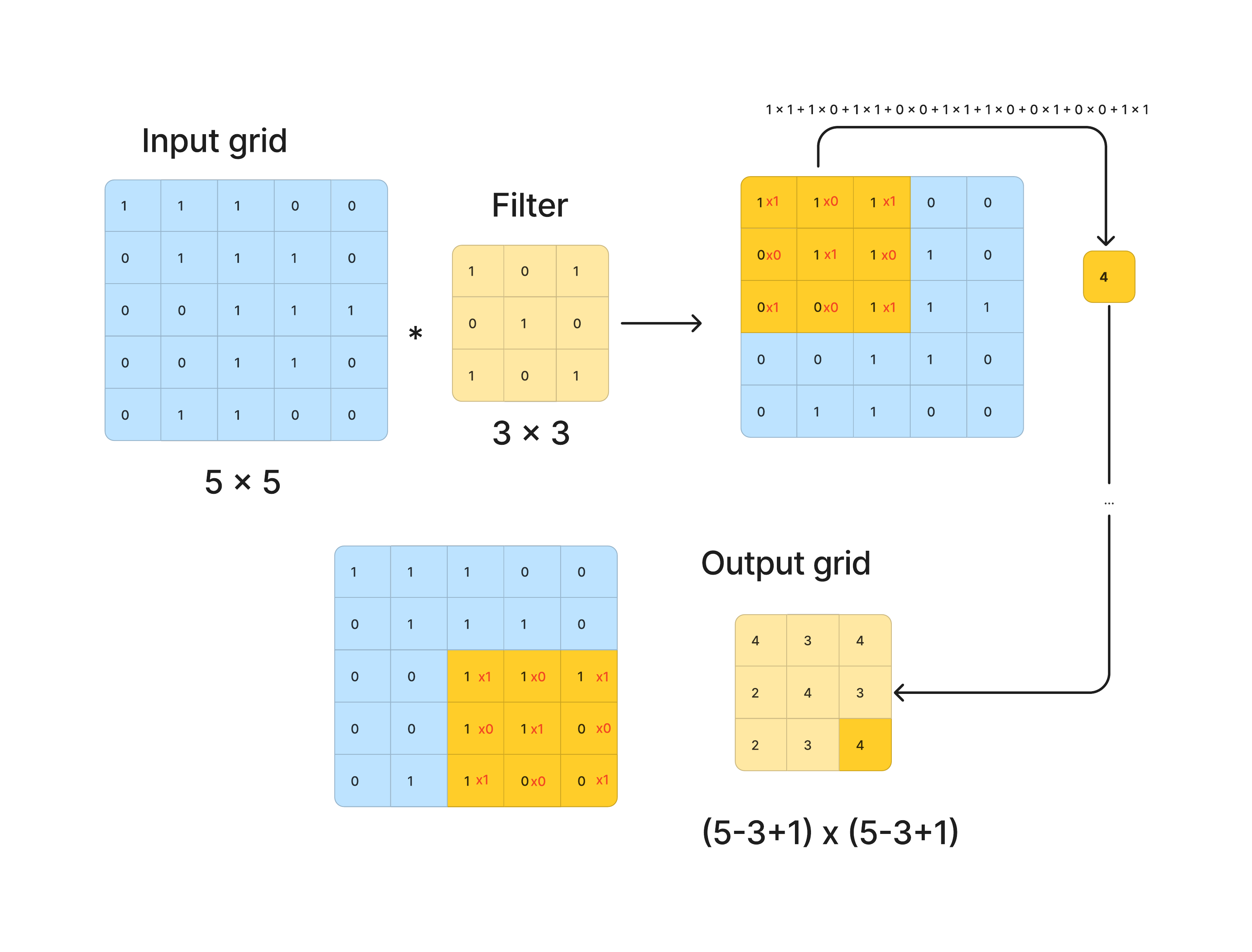}
    \caption{Schematic representation of a convolution between a $5\times5$ grid and a $3\times3$ filter, with depth 1. The filter is slid across the grid and the dot-product is performed at each alignment (represented by the yellow subset of the grid in the figure), until all the grid is covered and the $3\times3$ output grid is completed.}
    \label{fig:convolution}
\end{figure}

\begin{figure}
    \centering
    \includegraphics[width=0.5\textwidth]{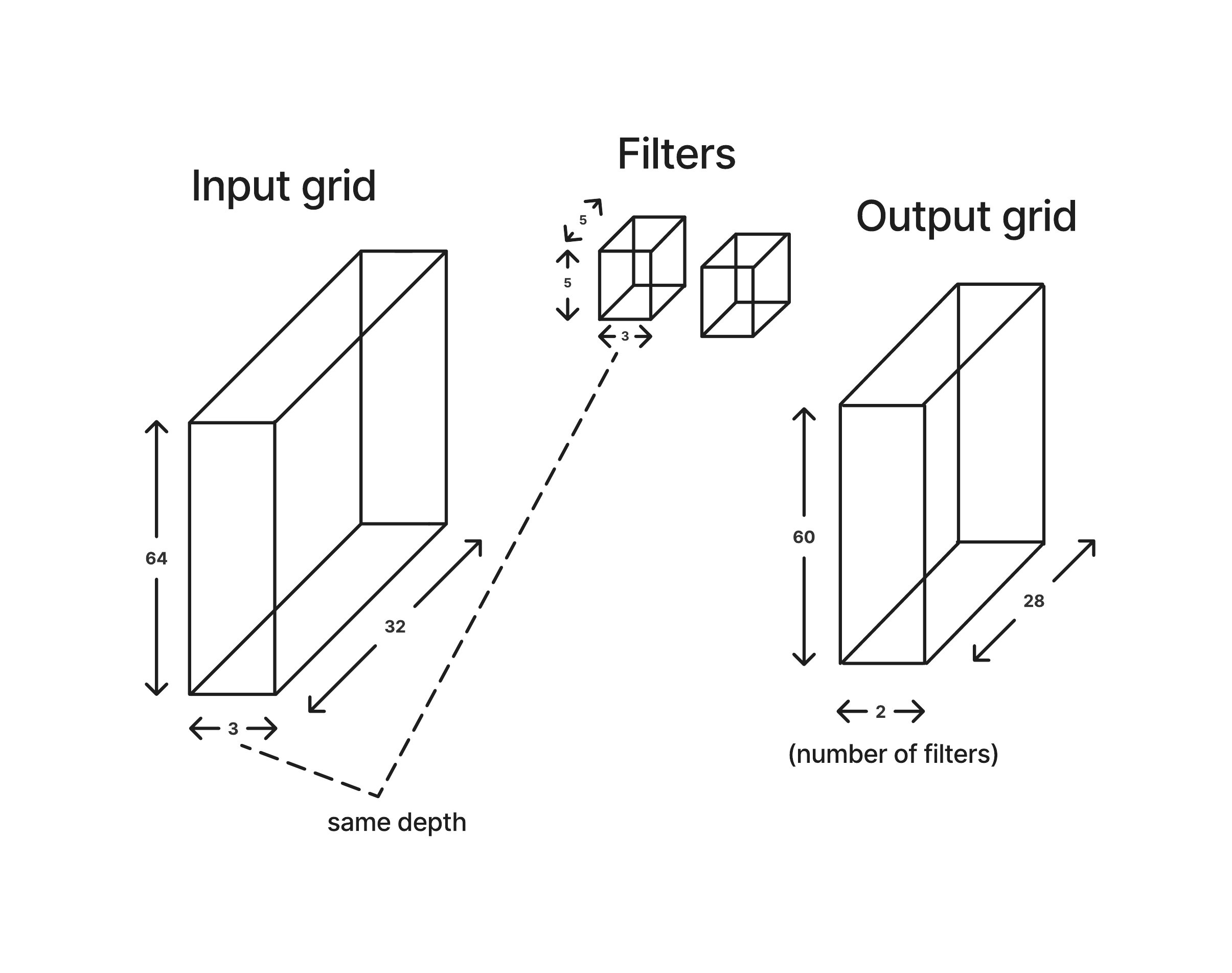}
    \caption{The convolution between a $64\times32\times3$ input grid with two $5\times5\times3$ filters gives a $60\times28\times2$ grid. Note that the depth of the filter and that of the grid must be the same, while the depth of the output grid is given by the number of filters.}
    \label{fig:input-output}
\end{figure}

The convolution operation is performed by shifting the filter across each possible location in the grid, performing a dot-product of its elements with those of the overlapping portion of the grid. To illustrate this, let us consider the example shown in Fig. \ref{fig:convolution}: here $L=B=5$, $F=3$ and $d=1$, for simplicity; the filter is multiplied element-by-element with the first $3\times3$ portion of the grid (starting from top left) and the products are summed, giving the first entry of the output map 
$ 1\times1+1\times1+1\times1+1\times1 = 4$, where multiplications with zeros are omitted, then the filter is shifted by one position in horizontal and the same process is repeated to obtain the second entry of the ouput map, until all the possible shifts of the filter on the image are performed. The number of possible positions to place the filter is defined by the number of the possible alignments along the height ($L-F+1$) times the number of alignments along the width ($B-F+1$), $3 \times 3$ in the example. This kind of operation guarantees that each of the output values (\emph{features}) of this operation are not connected indiscriminately to all the elements of the input data, but only to a local region, in order to recognize shapes and patterns. This is suited to map a local overdensity in the mass density map of a simulation to a substructure, in our application of CNNs for subhalo finding. 
Fig. \ref{fig:input-output} represents a more complex configuration, with $L=64$, $B=32$, $d=3$ and $F=5$. The convolution works like in the example with $d=1$, the only difference being that the dot product at each position of the grid is performed across the third dimension as well, thus outputting a $(64-5+1) \times (32-5+1)$ grid. Then, the depth of the output grid is built by stacking the output of convolutions with different filters (2 in the example). Here we are not considering \emph{strides} larger than 1, i.e. the possibility of shifting the filter by more than one positions at each step. For a complete treatment we refer to Chapter 8 of \citet{NN}.  \\
In general, for the $q$-th layer of a CNN having as input a grid with dimensions $L_q \times B_q \times d_q$, the depth $d_q$ is given by the number of filters applied in the $(q-1)-th$ layer. The number and dimensions of filters define the number of parameter, also known as the \emph{capacity} of the model. The underlying idea is that each filter is capable of capturing a particular spatial pattern, so a large number of them is required to capture all the patterns in the input image.\\
If one describes the parameters of the $p$-th filter in the $q$-th layer with the tensor $w_{ijk}^{(p,q)}$, with $i,j,k$ referring to the positions along the height, width and depth of the filter respectively, and the $k$-th \emph{feature map} (i.e. the output of a convolution operation) in the $q$-th layer as $h_{ijk}^{(q)}$, then the $p$-th feature map of the next layer is obtained as:
\begin{equation}
    \begin{aligned}
    h_{ijp}^{(q+1)} = \sum_{r=1}^{F_q} \sum_{s=1}^{F_q} \sum_{k=1}^{F_q} w_{rsk}^{(p,q)} h_{i+r-1,j+s-1,k}^{(q)} + b^{(p,q)} \\ \forall i \in [1,L_q-F_q+1], \; \forall j \in [1,B_q-F_q+1], \; \forall p \in [1,d_{q+1}];
    \end{aligned}
\end{equation}
where $b^{(p,q)}$ is the \emph{bias} for the $p$-th filter of the $q$-th layer, an extra parameter which is summed to the element-by-element product during to the convolution, thus only producing a uniform offset in the feature map, and $d_{q+1}$ is equal to the number of filters in the $q$-th layer.

Typically, in multilayer architectures, the filters in early layers detect more primitive shapes, whereas the filters in later layers detect more complex patterns, given by the composition of the more primitive shapes. This can be expressed more formally by saying that a convolution in the $q$-th layer increases the \emph{receptive field} of a feature, which captures information on larger scales if compared to the features of the earlier layers. For the example shown in Fig. \ref{fig:convolution}, once a $ 3 \times 3$ filter is applied, each feature contains information coming from $3\times3$ pixels of the previous layer and if one applies again the filter, being the layer of the same size as the filter, one gets only one feature which now contains information coming from a larger scale, corresponding to all the $5\times5$ pixels of two layers before.
In addition to convolution layers, i.e. layers connected to the next ones by a convolution operation, there are other two kind of layers typically present in the architecture of a CNN, which are called {\em pooling} and {\em ReLU} (Rectified Linear Unit) layers.

\subsection{Rectified Linear Unit and Pooling}
The Rectified Linear Unit (ReLU) is a function defined as:
\begin{equation}
    f_{ReLU} (x) = max \{ x,0 \}.
\end{equation}
This is very commonly used in Artificial Neural Network, especially CNNs. It is applied to (intermediate) outputs in order to introduce nonlinearity in the model one wants to approximate through the network, since the operation of convolution maps a region of the input into a feature map through a linear operation. In a CNN, it is applied to each of the $L \times B \times d$ values of a given layer, thus it does not change the layer size and it usually follows a convolution operation.

Pooling is an operation that acts on small regions of each layer and produces layers of the same depth, giving back for each region the maximum value; this is known as max pooling, which is way more diffused than other methods, like average pooling. Pooling, like ReLU, introduces nonlinearity in the model and also increases the size of the receptive field of a pixel.

\end{appendix}

\end{document}